\documentclass[12pt]{iopart}
\pdfoutput=1
\usepackage{color}
\usepackage{amssymb}

\usepackage{graphicx} 

\def\beqra{\begin{eqnarray}}
\def\eeqra{\end{eqnarray}}
\def\beq{\begin{equation}}
\def\eeq{\end{equation}}
\def\etap{\eta^\prime}
\def\etain{\eta_{in}}
\def\ds{\displaystyle}

\def\vp{\varphi}

\def\bx{{\bf{x}}}
\def\bk{{\bf{k}}}
\def\bp{{\bf{p}}}
\def\bq{{\bf{q}}}

\def\bv{{\bf{v}}}
\def\bw{{\bf{w}}}
\def\bl{{\bf{l}}}
\def\bV0{{\bf{V_0}}}
\def\re#1{(\ref{#1})}

\def\alt{\stackrel{<}{\sim}}
\begin{document}

\begin{flushright}
{\small UMN-TH-3134/13}
\end{flushright}
\title{
Galilean invariance and the consistency relation for the nonlinear  squeezed bispectrum of large scale structure
}
\author{Marco Peloso$^{1,2}$}
\ead{peloso@physics.umn.edu}
\author{Massimo Pietroni$^2$}
\ead{pietroni@pd.infn.it}
\vskip 0.3 cm
\address{
$^1$School of Physics and Astronomy, University of Minnesota, Minneapolis, 55455, USA\\
$^2$INFN, Sezione di Padova, via Marzolo 8, I-35131, Padova, Italy}

\begin{abstract}
We discuss the constraints imposed on the nonlinear evolution of the Large Scale Structure (LSS) of the universe by galilean invariance, the symmetry relevant on subhorizon scales. Using Ward identities associated to the invariance, we derive fully nonlinear consistency relations between  statistical correlators of the density and velocity perturbations, such as the power spectrum and the bispectrum. These relations are valid up to $O (f_{NL}^2)$ corrections. We then show that most of the semi-analytic methods proposed so far to resum the perturbative expansion of the LSS dynamics fail to fulfill the constraints imposed by galilean invariance, and are therefore susceptible to non-physical infrared effects. Finally, we identify and discuss a nonperturbative semi-analytical scheme which is manifestly galilean invariant at any order of its expansion.
\end{abstract}

\maketitle

\section{Introduction}
The Large Scale Structure of the Universe (LSS) provides an unique arena to test our understanding of the evolution of the Universe in recent cosmological epochs.  The main obstacle towards an accurate connection between theory and observations is represented by the many different sources of nonlinearity:  the nonlinear evolution of the dark matter (DM) fluid, redshift space distortions, halo and galaxy bias, etc.

Cosmological perturbation theory has received considerable attention in the recent literature as a promising tool to approach the nonlinear scales \footnote{ In this paper, we will consider only the first source of nonlinearity mentioned above, namely the treatment of nonlinearites in the evolution of the pure DM fluid in real space.}.  Being based on analytic, or semi-analytic, techniques, its computational times are in general greatly reduced with respect to those for N-body simulations. Moreover, although the basic formalism is derived for an Einstein-de Sitter cosmology, its extension to $\Lambda$CDM is straightforward, and also its formulations in non-standard cosmologies are feasible and under control. 

The main limitation of standard perturbation theory (PT) \cite{PT} is in the range of scales in which it can provide accurate results. As it was shown in \cite{JK06}, percent accuracy in the DM power spectrum (PS) can be attained for wave vectors $k\alt 0.2\, {\mathrm{h Mpc}^{-1}}$ for redshift $z\ge 1$, rapidly degrading at high $k$ and smaller $z$. Therefore, it cannot provide accurate predictions for the PS in the baryon acoustic oscillation (BAO) range of scales ($0.05\,  {\mathrm{h Mpc}^{-1}} \alt k \alt 0.25  {\mathrm{h Mpc}^{-1}}$) at low redshifts.  However, starting from the works of  Crocce and Scoccimarro \cite{RPTa,RPTb}, it has been realized that the PT expansion can be reorganized very efficiently, so that all-order PT contributions can be taken into account at lowest orders in the new expansion schemes \cite{ MP07b, RPTBAO, Taruya2007, Matsubara07, 2011MNRAS.416.1703E, Pietroni08, BV08, Bernardeau:2008fa, Anselmi:2010fs, Wang:2011fj, Wang:2012fr}.

Progress along these lines of research has been quite impressive: independent approaches are now able to produce, in a $O(1\;{\mathrm{min.}})$ time, DM PS which agree with those from high precision N-body simulations (and among themselves) at the percent level in the full BAO range of scales down to $z=0$ \cite{Valageas:2010yw, Anselmi:2012cn, Crocce:2012fa, Taruya:2012ut}, and to a few percent at smaller scales, up to $k \alt 0.8 \,{\mathrm{h Mpc}^{-1}}$ \cite{Anselmi:2012cn}. 

The purpose of this paper is to fully exploit the symmetry of the system, namely galilean invariance (GI), as a portal towards the nonlinear scales. The use of symmetry principles as tools to constrain the perturbative and nonperturbative sectors is common practice in quantum field theory. The consequences of the fundamental symmetries are enforced by Ward identities and consistency conditions, and provide powerful constraints on the structure of the fully renormalized (or effective) field theory. These constraints must be satisfied, besides perturbation theory,  by any other viable approximation scheme. 

GI in the context of the LSS was discussed in \cite{Scoccimarro:1995if}. Since a galillean transformation (GT) corresponds to a velocity perturbation of infinite wavelength, the link between GI and the infrared (IR) sector of the PT expansion was elucidated. In particular, the fact that leading IR divergences (emerging for a scale-free PS $P\sim k^n$ with $n\le -1$) cancel out at any order in PT \cite{Jain:1995kx}, was clearly understood as a consequence of GI. This fact was then used to identify which modifications of the full Euler-Poisson system respect GI and are therefore free of IR divergences, selecting the Zel'dovich dynamics as the only viable approximation.

In this paper we build upon this previous work, with a different focus. Namely, we stick to the exact dynamics, described in terms of the Vlasov equation and the Euler-Poisson
system, and:
\begin{itemize}
\item we derive the Ward identities, or consistency relations, enforced by GI on the correlators, {\it at the fully nonlinear level}. An example of it is the relation between the fully non-linear bispectrum and the fully non-linear PS of the density contrast $\delta$ given by  eq.~\re{P-B-finalid}: 
\begin{eqnarray}
&&  \!\!\!\!\!\!\!\!\!\!\!\!\!\ \!\!\!\!\!\!\!\!\!\!\!\!\!\  \!\!\!\!\!\!\!\!\!\!\!\!\!\ 
\lim_{k \to 0} B_\delta \left( k , q , \vert \bq + \bk \vert ; \tau , \tau' , \tau'' \right) \nonumber\\
&&  \!\!\!\!\!\!\!\!\!\!\!\!\!\ \!\!\!\!\!\!\!\!\!\!\!\!\!\  \!\!\!\!\!\!\!\!\!\!\!\!\!\ 
= \left[ -  \, \frac{\bq \cdot \bk}{k^2} 
\frac{D_+ \left( \tau' \right) -  D_+ \left( \tau'' \right) }{D_+ \left( \tau \right)}
+  \frac{6 f_{\rm NL}  \Omega_{m,0} H_0^2  }{k^2 T \left( k \right) } 
\frac{D_+ \left( \tau_{\rm 0} \right)  }{D_+ \left( \tau \right)}
\right] P_\delta (k; \tau, \tau)  P_\delta (q;\tau',\tau'') \nonumber\\
&&  \!\!\!\!\!\!\!\!\!\!\!\!\!\ \!\!\!\!\!\!\!\!\!\!\!\!\!\  \!\!\!\!\!\!\!\!\!\!\!\!\!\ + {\rm O } \left( k^0 , f_{\rm NL}^2 \right) \,, \nonumber\\
 \end{eqnarray}
(see Sects. \ref{galilean-tree} and \ref{fNL} for the  definitions of the various quantities). While it is straightforward to verify this relation at the tree level, Ward identities guarantee that this relation holds at the full non-linear level. This is completely analogous to the 
non-renormalization of the electric charge in QED through the nonperturbative $Z_1=Z_2$ identity. Finally, we stress that this relation also accounts for the  present  dark energy;

\item we show that the nonperturbative semi analytic approaches mentioned above are generally not well-behaved from the GI point of view, and that, indeed, most of them are plagued by spurious IR effects;
\item we identify and discuss a galilean invariant PT resummation scheme, the `eikonal renormalized perturbation theory' (eRPT), introduced in \cite{Anselmi:2012cn}.
\end{itemize}

The paper is organized as follows. In Sect. \ref{galilean-tree}, we introduce GI in physical and comoving coordinates, where it assumes the form of a time-dependent velocity boost. Then we show that both the Vlasov equation and the continuity and Euler equations derived from it are GI. In Sect.~\ref{PIn} we review the path-integral approach of \cite{MP07b} as an useful formalism to discuss symmetry properties at the fully nonperturbative level. In particular, we show how to extend this formalism from the center of mass (c.o.m.) frame to any frame related to it by a GT. Then, in Sect.~\ref{Ward} we show how to derive the Ward identities of GI, and we focus on those linking the two- and three-point correlators. In Sect.~\ref{fNL}, we discuss how this relation is affected by an initial nongaussianity, leading to the result written above. In Sect.~\ref{PTG} we discuss the link between GI and the IR sector of the loop corrections. We show how a GT can be implemented in a diagrammatic language and, finally, introduce a procedure to check if a given approximation scheme respects the GI constraints on the IR sectors and is therefore free from spurious IR effects. The link between GI and IR sensitivity is further elaborated in Sect.~\ref{GTIR}.  In Sect.~\ref{resummations} we consider the PT resummation schemes and show that most of them are not free from spurious IR effects. On the other hand we show that the eRPT \cite{Anselmi:2012cn} respects both our test procedure and the Ward identities at any order in the expansion. Finally, in Sect.~\ref{conclus}, we discuss the  implications of our results.

\section{Galilean invariance of the tree level action}
\label{galilean-tree}

A GT in physical coordinates reads:
\beqra
&& {\bf R}'={\bf R} - \bw \,t \nonumber\\
&& {\bf V}'={\bf V} - \bw \,.
\label{gp}
\eeqra
 Comoving coordinates and conformal time are defined, respectively, as
\beqra
&&\bx = \frac{{\bf R}}{a}\,,\nonumber\\
&&d \tau =\frac{d t}{a}\,,
\eeqra
where $a(\tau)$ is the scale factor. Moreover, the peculiar velocity and its canonically conjugated momentum are $\bv = d \bx/d \tau$ and $\bp = a m\,\bv$. 
The transformations \re{gp} then take the form
\beqra
&& \bx'=\bx - \bw\, T \,,\nonumber\\
&& \bv'=\bv  - \bw \,\dot T\,,
\label{gc}
\eeqra
where 
\beqra
&& T (\tau)= \frac{1}{a(\tau)}\int_0^\tau d\tau' a(\tau')\,,\nonumber \\
&& \dot{T}(\tau) = \frac{d T}{d \tau} = 1 - {\cal H} T \;\;\;\;\;\;\left( {\cal H}\equiv\frac{1}{a}\frac{d a}{d\tau} \right)\,,
\label{TTdot}
\eeqra
that is, in comoving coordinates and conformal time, GT can be seen as a boost  by a {\em time-dependent} velocity $\tilde{\bw}(\tau) \equiv \bw\, \dot{T}(\tau)$. This makes the consequences of GT on multi-point correlators  non-trivial.

We note that, under the transformation \re{gc}, the peculiar momentum transforms as $\bp'=\bp - a m \, \dot T \,\bw$. 

The particle distribution function is a GT-invariant scalar: $f'(\bx',\bp',\tau)=f(\bx,\bp,\tau) = f(\bx' + \bw\, T ,\bp'+ a m \, \dot T \,\bw,\tau)$, and the total time-derivative operator,
\beq
\frac{d\;}{d\tau}\equiv \frac{\partial\;}{\partial \tau} + v^i \frac{\partial\;}{\partial x^i} + \dot{p}^i\frac{\partial\;}{\partial p^i}\,,
\eeq
is GT-invariant. Therefore, the Vlasov equation, namely, 
\beq
\frac{d\;}{d\tau} f(\bx,\bp,\tau) = \frac{d\;}{d\tau}f'(\bx',\bp',\tau)\,,
\eeq
is GT-invariant.
Indeed,
\beqra
&& \big(\frac{\partial\;}{\partial \tau} + {v^i}' \frac{\partial\;}{\partial {x^i}'} + \dot{p^i}'\frac{\partial\;}{\partial {p^i}'}\big)f'(\bx',\bp',\tau) \nonumber\\
&&= \big(\frac{\partial\;}{\partial \tau} + ({v^i}-w^i \dot T) \frac{\partial\;}{\partial {x^i}} +(  \dot{p^i} -\frac{d\;}{d\tau}(a m \, \dot T) w^i)\frac{\partial\;}{\partial {p^i}}\nonumber\\
&& \qquad +w^i \dot T \frac{\partial\;}{\partial {x^i}'} +\frac{d\;}{d\tau}(a m \, \dot T) w^i)\frac{\partial\;}{\partial {p^i}'} \big)  f(\bx' + \bw\, T ,\bp'+ a m \, \dot T \,\bw,\tau) \nonumber\\
&&= \big(\frac{\partial\;}{\partial \tau} + {v^i} \frac{\partial\;}{\partial {x^i}} + \dot{p^i}\frac{\partial\;}{\partial {p^i}}\big)f(\bx,\bp,\tau) \,.
\label{Vlasov-invariance}
\eeqra

The Vlasov equation can be cast in a series of equations for the moments of the distribution function
\cite{PT}
\begin{eqnarray}
&&
\int d^3 \bp f \left( \bx , \bp , \tau \right) \equiv \rho \left( \bx , \tau \right) \equiv \rho_0 \left( t \right) \left[ 1 + \delta \left( \bx , \tau \right) \right] \nonumber\\
&&
\int d^3 \bp \frac{p_i}{a m} \,  f \left( \bx , \bp , \tau \right) \equiv \rho \left( \bx , \tau \right) \, v_i \left( \vec{x} ,\, \tau \right)
 \nonumber\\
&&
\int d^3 \bp \frac{p_i p_j}{a^2 m^2} \,  f \left( \bx , \bp , \tau \right) \equiv        
 \rho \left( \bx , \tau \right) \, v_i \left( \bx ,\, \tau \right) \, v_j \left( \bx ,\, \tau \right)
+ \sigma_{ij} \left( \bx ,\, \tau \right) \nonumber\\
&& \dots
\end{eqnarray}
The series can be consistently truncated to the continuity and Euler equations by setting $\sigma_{ij} = 0$ (this truncation being known as the ``single stream approximation''):
\begin{eqnarray}
&&  \left[  \frac{\partial }{\partial \tau} +  \bv \cdot \nabla \right] \delta = 
- \left( 1 + \delta \right) \nabla \bv \, ,  \nonumber\\
&& \left[  \frac{\partial }{\partial \tau} +  \bv \cdot \nabla \right] \bv = - {\cal H } \bv  - \nabla \phi \, , 
\label{cont-Eul}
\end{eqnarray}
where $\phi$ is the gravitational potential, which, on super-horizon scales, obeys the Poisson equation
\begin{equation}
\nabla^2 \phi = \frac{3}{2} {\cal H}^2\; \Omega_m \delta \,,
\label{poisson}
\end{equation}
where $\Omega_m(a)$ is the time dependent matter density parameter.
Throughout this paper we will assume a $\Lambda$CDM background cosmology, or, more generally, a dynamical dark energy scenario in which the dark energy fluctuations can be neglected on all scales of interest.

The expressions (\ref{cont-Eul}) have been written in such a way that both LHS's and RHS's are invariant under a GT. We note that on the LHS's both the linear and nonlinear terms transform nontrivially, but that the two nontrivial pieces in the transformation cancel each other, so that the total operator is invariant
(the cancellation is inherited from that of the Vlasov equation shown in (\ref{Vlasov-invariance})). We also note that the gravitational potential transforms nontrivially under a GT \cite{Scoccimarro:1995if}.

This system can be reformulated with a compact notation  \cite{RPTa,RPTb,MP07b} by Fourier transforming 
\begin{equation}
\delta \left( \bx , \tau \right) = \int d^3 \bp\; {\rm e}^{i \bx \cdot \bk } \, \delta \left( \bk , \tau \right) \, , 
\end{equation}
and analogously for $\bv$, by assuming no vorticity,
\begin{equation}
\bv \left( \bk , \tau \right) = - i \frac{\bk}{k^2} \theta \left( \bf k , \tau \right) \, , 
\label{v-theta}
\end{equation}
with $k^2=\bk\cdot\bk$, and by introducing the doublet
\begin{equation}
\left( \begin{array}{c} 
\varphi_1 \left( \bk , \eta \right) \\
\varphi_2 \left( \bk , \eta \right) 
\end{array} \right) \equiv
{\rm e}^{-\eta} \left( \begin{array}{c}
\delta \left( \bk , \eta \right) \\ - \theta \left( \bk , \eta \right) / {\cal H} f 
\end{array} \right)
\;\;\;,\;\;\; \eta \equiv \, {\rm ln } \,  \frac{D_+(\tau)}{D_+(\tau_{in})} \, .
\label{doublet}
\end{equation}
In the definitions above, we have introduced the new time variable, $\eta$, in terms of the linear growth factor $D_+$, and we have also used the linear growth function $f(\eta)\equiv 1/{\cal H} \;d \eta / d\tau $.

With this notation, eqs. (\ref{cont-Eul}) can be reformulated as
\begin{equation}
\!\!\!\!\!\!\!\!\!  \!\!\!  \left( \delta_{ab} \partial_\eta + \Omega_{ab} \right) \varphi_b \left( \bk , \eta \right) = {\rm e}^\eta 
\int d^3 \bp \, d^3 \bq \,  \gamma_{abc} \left( \bk ,\, - \bp ,\, - \bq \right) \varphi_b \left( \bp ,\, \eta \right) 
 \varphi_c \left( \bq ,\, \eta \right) \, , 
\label{compact}
\end{equation}
where
\begin{equation}
\ds {\bf \Omega} = \left( \begin{array}{cc} 
1 & - 1 \\ - \frac{3}{2} \frac{\Omega_m}{f^2} &  \frac{3}{2} \frac{\Omega_m}{f^2} 
\end{array} \right)\,,
\label{BIGOMEGA}
\end{equation}
and  where the only non vanishing components of the vertex function are
\begin{eqnarray}
&&
\!\!\!\!\!\!\!\!\!\gamma_{121} \left( \bk ,\, \bp ,\, \bq \right) = \frac{1}{2} \delta_D \left( \bk + \bp + \bq \right) \alpha \left( \bp , \bq \right) \;\;,\;\;
\!\!\!\!\!\!\!\!\!\gamma_{121} \left( \bk , \bp , \bq \right) = \gamma_{112} \left( \bk , \bq , \bp \right)  \;\;, \nonumber\\
&&
\!\!\!\!\!\!\!\!\!\gamma_{222} \left( \bk ,\, \bp ,\, \bq \right) =  \delta_D \left( \bk + \bp + \bq \right) \beta \left( \bp , \bq \right) 
\, ,
\end{eqnarray}
with $\delta_D$ being the Dirac $\delta-$function, and
\begin{equation}
\alpha \left( \bp, \bq \right) = \frac{\left( \bp+\bq \right) \cdot \bp}{p^2} \;\;,\;\;
\beta \left( \bp, \bq \right) = \frac{\left( \bp+\bq \right)^2 
\, \bp \cdot \bq  }{2 p^2 q^2} \, .
\label{al-be}
\end{equation}
Under a GT the field transforms as
\begin{equation}
\varphi_a \left( \bk , \eta \right) \rightarrow {\rm e}^{i \bk \cdot \bw \, T \left( \eta \right)}
\varphi_a \left( \bk , \eta \right) + i  \,  \bk \cdot \bw \, {\rm e}^{-\eta}  \partial_\eta T \left( \eta \right) \, \delta_D \left( \bk \right)\,\delta_{a2} \, .
\label{GT-phi}
\end{equation}
The second piece must be retained in the GT of the right hand side of (\ref{compact}), as the vertex functions have a pole when their second or third argument vanishes. Their presence is actually crucial to ensure the covariance of  (\ref{compact}). Indeed, under a GT, eq.  (\ref{compact}) changes into
\begin{eqnarray}
& & \!\!\!\!\!\!\!\! \!\!\!\!\!\!\!\! \!\!\!\!\!\!\!\! \!\!\!\!\!\!\!\! \!\!\!
\left( \delta_{ab} \partial_\eta + \Omega_{ab} \right) \left[  {\rm e}^{i \bk \cdot \bw \, T \left( \eta \right)} \varphi_b \left( \bk , \eta \right) \right] 
=   {\rm e}^\eta 
\int d^3 \bp \, d^3 \bq \,  \gamma_{abc} \left( \bk ,\, - \bp ,\, - \bq \right) {\rm e}^{i \left( \bp + \bq \right) \cdot \bw \, T \left( \eta \right)} \times
 \nonumber\\
& & 
 \left[ \varphi_b \left( \bp , \eta \right) +  i \bp \cdot \bw {\rm e}^{-\eta} \partial_\eta T \delta_D \left( \bp \right) \delta_{b2} \right] \left[ \varphi_c \left( \bq , \eta \right) +  i \bq \cdot \bw {\rm e}^{-\eta} \partial_\eta T \delta_D \left( \bq \right)\delta_{c2} \right] \, .  \nonumber\\
\label{compact-transf1}
\end{eqnarray}
There are four terms at right hand side. The one obtained from the two second pieces in the square parenthesis vanishes, since the various $\delta-$functions force $\bk=0$, and the vertex function vanishes when its first argument vanishes, due to translation invariance. The two ``mix-terms'' instead add up to cancel the derivative of the phase on the left hand side.
Indeed, using
\begin{equation}
\bp\cdot\bw  \, \delta_D \left( \bp \right) \, \gamma_{abc} \left( \bk , \bp , \bq \right)  = \frac{1}{2}
\frac{\bp \cdot \bq\,\bp \cdot \bw}{p^2}\delta_{b2} \delta_{ac} \delta_D \left( \bk + \bp + \bq \right) \delta_D \left( \bp \right) \, ,
\label{limitg}
\end{equation}
eq. (\ref{compact-transf1}) rewrites
\begin{eqnarray}
& & 
 {\rm e}^{i \bk \cdot \bw \, T \left( \eta \right)} \left[  \left( \delta_{ab} \partial_\eta +  \Omega_{ab} \right)  \varphi_b \left( \bk , \eta \right)  +     i \bk \cdot \bw \partial_\eta T  \varphi_a \left( \bk , \eta \right) \right] \nonumber\\
& &  \;\;\;\;\;\;\;\;\;\;\;\;\;\;\;\;\;\;
=   {\rm e}^{i \bk \cdot \bw \, T \left( \eta \right)}  \Bigg[ {\rm e}^\eta
 \int d^3 \bp d^3 \bq \gamma_{abc} \left( \bk , - \bp , - \bq \right)   \varphi_b \left( \bp , \eta \right)  \varphi_c \left( \bq , \eta \right)\nonumber\\
 && \;\;\;\;\;\;\;\;\;\;\;\;\;\;\;\;\;\;  \;\;\;\;\;\;\;\;\;\;\;\;\;\;\;\;\;\;+ \partial_\eta T \,  \varphi_a \left( \bk , \eta \right) 
   \int d^3 \bp \frac{i \bp \cdot \bw \, \bp \cdot \bk}{p^2} \delta_D \left( \bp \right)  \Bigg] \, .
\nonumber\\
\label{compact-transf2}
\end{eqnarray}
We indeed see that the last term at the LHS cancel against the last term at the RHS, and 
eq. \re{compact} is recovered also in the new frame. We stress that the two terms that cancel each other come from the transformation of the linear and the nonlinear part of \re{compact}.

As done in \cite{MP07b} we introduce the action
\begin{eqnarray}
&& S= S_{\rm free} + S_{\rm int} \nonumber\\
&&\;\;\;= \int d \eta d \etap  \,d^3 \bk \,\chi_a \left( - \bk , \eta \right) \, g_{ab}^{-1}(\eta,\etap)\, \varphi_b \left( \bk , \eta \right)\nonumber\\
&& \;\;\;- \int d \eta d^3 \bk  d^3 \bp \, d^3 \bq  {\rm e}^\eta  \,  \gamma_{abc} \left( - \bk ,\, - \bp ,\, - \bq \right) 
 \chi_a \left(  \bk , \eta \right)  \varphi_b \left( \bp ,\, \eta \right) 
 \varphi_c \left( \bq ,\, \eta \right),
\label{action}
 \end{eqnarray}
where
\beq
 g_{ab}^{-1}(\eta,\etap) = \delta_D(\eta-\etap) \left( \delta_{ab} \partial_{\etap} + \Omega_{ab} \right)\,,
\label{gil}
 \eeq
is the inverse linear propagator, which, inverted imposing casual initial conditions, gives the linear propagator,
\beq
g_{ab}(\eta) = \left(B_{ab} + e^{-5/2 \eta} A_{ab}\right) \, \theta(\eta)\,,
\eeq
with $\theta(x)$ the Heaviside step function, and
\beq
 {\bf B} = \frac{1}{5}\left( \begin{array}{cc} 
3 & 2 \\3& 2
\end{array} \right)\,,\;\;\;\;\;\;  {\bf A} = \frac{1}{5}\left( \begin{array}{cc} 
2 & -2 \\3& -3
\end{array} \right)\,.
\eeq
In order to obtain the above explicit expression, we have approximated $\Omega_m/f^2 \simeq 1$ in  \re{BIGOMEGA}, see \cite{PT,Pietroni08} for a discussion of this approximation and for an assessment of its numerical validity at the non-linear level. 
This approximation allows for a simple expression of the linear propagator, but it is not used in the reminder of the paper, and therefore   it by no means affects the following discussion.

Extremizing the action in eq.~\re{action} with respect to $\chi_a$, gives the equation of motion \re{compact}. To enforce invariance under GT, 
we impose that, under a GT,
\begin{equation}
\chi_a \left( \bk , \eta \right) \rightarrow {\rm e}^{i \bk \cdot \bw \, T \left( \eta \right)}
\chi_a \left( \bk , \eta \right) \, ,
\label{GT-chi}
\end{equation}
so that, under (\ref{GT-phi}) and  (\ref{GT-chi}), 
\begin{equation}
S \rightarrow S + \int d \eta d^3 \bk \chi_a \left( -  \bk , \eta \right) \left( \delta_{a2} \partial_\eta + \Omega_{a2} \right) i \bk \cdot \bw {\rm e}^{-\eta} \partial_\eta T \delta_D \left( \bk \right) \, ,
\label{change-S}
\end{equation}
which is invariant up to a term that gives a vanishing contribution to the equation of motion \re{compact}. The invariance is the result of the same cancellation between the linear and the nonlinear term that we have already observed in (\ref{compact-transf2}), and that we have also commented after eq. (\ref{poisson}). 

\section{Path integral formulation and frame fixing}
\label{PIn}

Ref.  \cite{MP07b} provided a path integral formulation of the system (\ref{compact}). The formulation is characterized by the generating functional
\begin{eqnarray}
& & \!\!\!\!\!\!\! \!\!\!\!\!\!\! \!\!\!\!\!\!\! \!\!\!\!\!\!\!
Z \left[ J_a ,\, K_b ; \bV0 \right]  =  \int {\cal D} \varphi_a {\cal D} \chi_b \, {\rm exp } \, \Bigg\{ - \frac{1}{2} \int d^3 \bk  \,\chi_a \left( - \bk , 0 \right) P_{ab}^0 \left( k \right) \, \chi_b \left( \bk  , 0 \right) + i S \nonumber\\
& &  \!\!\!\!\!\!\! \quad\quad\quad + i \int d \eta d^3 \bk \left[ J_a \left( - \bk , \eta \right) \varphi_a \left( \bk , \eta \right) +  K_a \left( - \bk , \eta \right) \chi_a \left( \bk , \eta \right) \right] + i S_{\rm ff} \left[ \bV0 \right] \Bigg\} \, ,  \nonumber\\
\label{Z}
\end{eqnarray}
where $S$ is the action (\ref{action}).  Correlators with $n$ $\vp_a$ fields and $m$ $\chi_b$ ones are obtained, as usual, by taking $n$ derivatives of the path integral with respect to the source $i J_a$ and $m$ derivatives with respect to $iK_b$. 
In particular, second derivatives give the nonlinear two-point correlator, {\it i.e.} the PS, 
\beq
 \!\!\!\!\! \!\!\!\!\!\!\! \!\!\!\!\!\!\! \!\!\!\!\!\!\! \!\!\!\!\!\!\!\left.\frac{\delta W}{\delta J_a(-\bk,\eta)\delta J_b(-\bk',\eta') } \right|_{J_a=K_a=0} = i\langle \varphi_a \left( \bk , \eta \right)  \varphi_b \left( \bk' , \etap \right) \rangle_{\bV0}= i \delta_D(\bk+\bk') P_{ab}(k;\eta,\etap)\,,
 \label{PSdef}
\eeq
and the nonlinear propagator
\beq
 \!\!\!\!\! \!\!\!\!\!\!\! \!\!\!\!\!\!\! \!\!\!\!\!\!\! \!\!\!\!\!\!\!\left.\frac{\delta W}{\delta J_a(-\bk,\eta)\delta K_b(-\bk',\eta') } \right|_{J_a=K_a=0} =  i\langle \varphi_a \left( \bk , \eta \right)  \chi_b \left( \bk' , \etap \right) \rangle_{\bV0}=- \delta_D(\bk+\bk') G_{ab}(k;\eta,\etap)\,,
\eeq
where $W\equiv-i \log Z$.

The expression in eq.~\re{Z}  holds for gaussian initial conditions (at the ``time'' $\eta = 0$, chosen when all the relevant modes are linear), encoded by the linear PS
\begin{equation}
 P^0_{ab}(k) \equiv  P^0_{ab}(k;0,0)=  u_a u_b \,  P^0 \left( k \right) \, ,
 \label{Plin}
\end{equation}  
where in the last equality $u_a = \left( \begin{array}{c} 1 \\ 1 \end{array} \right)$ selects the growing mode. Nongaussian initial conditions can be taken into account by adding 
trinliear, quadrilinear, and so on, terms in $\chi_a$ for the initial bispectrum, trispectrum, etc., respectively, see eq.~\re{Z-fNL}.

 In \re{Z}  we  introduced also a term linear in $\chi_a$, which was absent in \cite{MP07b},  the ``frame-fixing'' term
\begin{equation}
S_{\rm ff} \left[ \bV0 \right] =  -   \int d^3 \bk d \eta d \eta'  \chi_a \left( - \bk ,  \eta \right) g_{ab}^{-1} \left( \eta , \eta' \right) {\bar \varphi}_b \left( \bk , \eta' , \bV0  \right) \, ,  \nonumber\\
\label{Sgf}
\end{equation}
where
\begin{equation}
{\bar \varphi}_a \left( \bk , \eta , \bV0 \right)  \equiv  - i \,\delta_{a2} {\rm e}^{-\eta} \, \bk\cdot \bV0  \,  \delta_D \left( \bk \right)  \,\partial_\eta T .
\label{phi2-bar}
\end{equation}
Adding the term $S_{\rm ff}$  to the action selects the inertial frame in which the average physical velocity of particles is given by $\bV0$, and, correspondingly, the average peculiar velocity is $\bV0 \dot T$, see eq. \re{gc}. Indeed, we first notice that
\beq
\langle \bv(\bx)\rangle = \frac{1}{\mathrm{Vol.}} \int d^3 x\, \bv(\bx) = i\, \frac{(2 \pi)^3}{\mathrm{Vol.}} \, {\cal{H}} f\,e^\eta\,\int d^3 k\,\delta_D(\bk)\, \frac{\bk}{k^2} \vp_2(\bk)\,,
\eeq
where we have used eqs.~\re{v-theta} and \re{doublet} and ``Vol.'' stands for the total volume on which the average is taken. Setting $\vp_2(\bk)= {\bar \varphi}_2 \left( \bk , \eta , \bV0 \right) $, gives $\langle \bv(\bx)\rangle = \bV0 \dot T$ (where we have used the relation ${\cal H} f \partial_\eta T = \dot T$).

Then, we show that the inclusion of the linear term in $\chi_a$, eq.~\re{Sgf}, at the exponent of the path integral forces the field $\vp_a(\bk,\eta)$ to have a non-vanishing expectation value, given by eq.~\re{phi2-bar}. Indeed, when $S_{\rm ff}=0$, translational invariance of the action and of the initial conditions ensure that the expectation values of the field vanish. Translational invariance of the dynamics is encoded in the property of the linear vertex $\lim_{k\to 0}\gamma_{abc}(\bk,\bq,\bp) = 0$.

To see how a non-vanishing expectation value emerges when $S_{\rm ff}\neq 0$ we transform it away by making a change of integration variables in \re{Z} corresponding to eqs.~\re{GT-phi} and $\re{GT-chi}$ with $\bw=-\bV0$.
It gives
\beq
 \!\!\!\!\!\!\!\! \!\!\!\!\!\!\!\! \!\!\!\!\!\!\!\! \!\!\!\!\!\!\!\!
Z \left[ J_a,  K_b; \bV0 \right] = e^{i \int d \eta d^3 \bk J_a \left( - \bk , \eta \right) {\bar \varphi}_a \left( \bk, \eta , \bV0 \right)} \;Z \left[  J_a\,e^{i \bk \bV0 T \left( \eta \right) } ,  K_b \,e^{i \bk \bV0 T \left( \eta \right) }; \bV0={\bf 0 } \right] \, ,
\label{id1}
\eeq
where the generating functional on the RHS is obtained from eq.~\re{Z} by setting $S_{\rm ff}=0$ and by multiplying the sources by the GT phases. The field expectation value is obtained by taking the first derivative with respect to the source, and then setting the sources to zero,
\beq
\langle \vp_a(\bk,\eta) \rangle_{\bV0} =-i\; \left.\frac{d \log Z  \left[ J_a,  K_b;  \bV0 \right] }{d J_a(-\bk,\eta)}\right|_{J_a=K_a=0}\,.
\eeq
The only non-vanishing contribution to the derivative of the RHS of \re{id1}  comes from the factorized exponential, which gives precisely
\beq
\langle \vp_a(\bk,\eta) \rangle_{\bV0} =  {\bar \varphi}_a \left( \bk , \eta , \bV0 \right)\,.
\eeq

By taking the appropriate derivatives of eq.~\re{id1} with respect to the sources, we can  obtain the transformation law for any correlator of $\vp$ and $\chi$ fields between the center of mass frame $(\bV0=0)$ and a generic one with $\bV0\neq 0$. It is given by
\begin{eqnarray}
& &\!\!\!\!\!\!\!\! \!\!\!\!\!\!\!\! \!\!\!\!\!\!\!\!
\langle \varphi_{a_1} \left( \bq_1 , \eta_1 \right) \dots  \varphi_{a_n} \left( \bq_n , \eta_n \right) 
 \chi_{a_{n+1}} \left( \bq_{n+1} , \eta_{n+1} \right) \dots  \chi_{a_{n+m}} \left( \bq_{n+m}  , \eta_{n+m} \right) 
\rangle_{\bV0}  \nonumber\\
& &
\!\!\!\!\!={\rm e}^{-i \bV0 \left[ T \left( \eta_1 \right) \bq_1 + \dots + T \left( \eta_{n+m} \right) \bq_{n+m} \right] }\nonumber\\
&&\!\!\!\!\!\times \langle  \varphi_{a_1} \left( \bq_1 , \eta_1 \right) \dots  \varphi_{a_n} \left( \bq_n , \eta_n \right) 
 \chi_{a_{n+1}} \left( \bq_{n+1} , \eta_{n+1} \right) \dots  \chi_{a_{n+m}} \left( \bq_{n+m}  , \eta_{n+m} \right) 
\rangle_{\bV0 = \bf 0} \,,\nonumber\\
\label{general-corr-GT}
\end{eqnarray}
where we have taken all momenta $\bq_i\neq0$.
Since translational invariance requires the correlator to be proportional to $\delta_D\left(\sum_i \bq_i\right)$, we see that equal-time correlators, such as the PS from eq.~\re{PSdef} with $\eta=\etap$, are invariant under a GT.

\section{ Ward identities}
\label{Ward}

In   Sect.~\ref{galilean-tree}  we have seen that the the covariance of  the equations of motions, eq.~\re{compact}, under GT, emerges as a consequence of a relation between the tree-level (linear) inverse propagator and the tree-level vertex in the limit in which one of its $\vp_a$-ends corresponds to a velocity field in the long wavelength limit
(see eqs.~\re{compact-transf1}, \re{limitg}). In this section, we investigate how the galilean invariance of the dynamics (and of the initial conditions) manifests itself at the fully nonlinear level. We will find relations between correlators holding nonpertubatively, and therefore valid beyond the range of scales in which usual perturbative (standard and renormalized) methods can be applied.  Using a field-theoretical language, these ``consistency relations'' are the Ward identities of GT.

Let us consider a GT  of infinitesimal parameter $ \bw$. Under this transformation, the fields $\varphi_a$ and $\chi_a$ transform as
the infinitesimal versions of \re{GT-phi} and \re{GT-chi},  respectively
\begin{eqnarray}
\delta \varphi_a \left( \bk , \eta \right) & = &  i \bk \cdot  \bw \, \left[ T \left( \eta \right) \varphi_a \left( \bk , \eta \right) + {\rm e}^{-\eta} \partial_\eta T \left( \eta \right) \delta_D \left( \bk \right) \delta_{a2} \right] \, , \nonumber\\
\delta \chi_a \left( \bk , \eta \right) & = &  i \bk \cdot  \bw \, T \left( \eta \right)  \chi_a \left( \bk , \eta \right)  \, ,
\end{eqnarray}
 while the  generating functional $Z$ remains invariant:
\begin{eqnarray}
& &  \!\!\!\!\!\!\!\!  \!\!\!\!\!\!\!\! \!\!\!\!\!\!\!\!  \!\!\!\!\!\!\!\!  
  \delta Z = \, 
\int   {\cal D} \varphi_a \,   {\cal D} \chi_b \, {\rm exp } \left\{ \dots \right\} \, 
  \int d \eta d^3 k \left( \bk \cdot \bw \right)\times\nonumber\\
  &&  \Bigg\{ T \left( \eta \right) \Bigg[ J_a \left( - \bk , \eta \right) \varphi_a \left( \bk , \eta \right) +  K_a \left( - \bk , \eta \right) \chi_a \left( \bk , \eta \right) \Bigg] \nonumber\\ 
  &&+ \Bigg[ J_2 \left( - \bk , \eta \right) + \chi_a \left( - \bk ,\, \eta \right) \left( \delta_{a2} \partial_\eta+ \Omega_{a2} \right) \Bigg] {\rm e}^{-\eta} \partial_\eta T \left( \eta \right) \delta_D \left( \bk \right)\Bigg\} = 0 \, ,
\label{dZ1}
\end{eqnarray}
where the ellipses in the exponent denotes the terms present in $Z = \int   {\cal D} \varphi_a \,   {\cal D} \chi_b \, {\rm exp } \left\{ \dots \right\} $.

The 1PI Green functions are obtained from functional derivatives of the effective action $\Gamma$, related to $W = -i \log Z$ by   a Legendre transform: 
\begin{equation}
\!\!\!\!\!\!\!\!  \!\!\!\!\!\!\!\!  \!\!\!\!\!\!\!  \!\!\!\!\!\!\!\!  \Gamma \left[ {\tilde \varphi}_a , {\tilde \chi}_b \right] \equiv W \left[ J_a ,\, K_b \right] - \int d \eta d^3 \bk \left[ J_a \left( - \bk , \eta \right) {\tilde \varphi}_a \left( \eta ,\, \bf k \right) +  K_a \left( - \bk , \eta \right) {\tilde \chi}_a \left( \eta ,\, \bf k \right) \right] \,,
\end{equation}
where the `classical' fields in presence of sources are
\begin{equation}
{\tilde \vp}_a \left( \bk , \eta \right) =  \frac{ \delta W }{ \delta J_a \left( - \bk , \eta \right) } \;\;,\;\;
{\tilde \chi}_a \left( \bk , \eta \right) =  \frac{ \delta W }{ \delta K_a \left( - \bk , \eta \right) } \, ,
\label{field-sources}
\end{equation}
and
\begin{equation}
J_a \left( \bk , \eta \right) = - \frac{ \delta \Gamma }{ \delta {\tilde \varphi}_a \left( - \bk , \eta \right) } \;\;,\;\;
K_a \left( \bk , \eta \right) = - \frac{ \delta \Gamma }{ \delta {\tilde \chi}_a \left( - \bk , \eta \right) } \, .
\label{JK-Gamma}
\end{equation}
In the following, we will omit the overtilde on the classical fields but it should be clear that they are (source-dependent) expectation values, not be confused with the stochastic fields we have considered up to now. 
Starting from (\ref{dZ1}), one can take out of the path integral all the  field independent quantities, and express the sources through \re{JK-Gamma}. One is left with~\footnote{We note that we have disregarded the infinitesimal transformation of $S_{\rm ff}$ in eq. \re{dZ1}. Using  eq.~\re{id1}, we can see that the effect of this term is that, in a generic frame, all the $J$ and $K$ appearing in equations  \re{dZ1} to \re{JK-Gamma} should be multiplied by the phase  $e^{i \bk \bV0 T }$. The expression \re{dZ2} and the following ones are then unchanged.}
\begin{eqnarray}
& & 
  \int d \eta d^3 \bk \left( \bk \cdot \bw \right) \Bigg\{ T \left( \eta \right) \left[ 
    \frac{\delta \Gamma}{\delta \varphi_a \left( \bk  , \eta \right)} \varphi_a \left( \bk , \eta \right) +      \frac{\delta \Gamma}{\delta \chi_a \left( \bk  , \eta \right)} \chi_a \left( \bk , \eta \right) \right] \nonumber\\
& & + \left[ \frac{\delta \Gamma}{\delta \varphi_2 \left( \bk , \eta \right)} - \chi_a \left( - \bk ,\, \eta \right) \left( \delta_{a2} + \Omega_{a2} \right) \right] {\rm e}^{-\eta} \partial_\eta T \left( \eta \right) \delta_D \left( \bk \right) \Bigg\} = 0 \, .
\label{dZ2}
\end{eqnarray}

Different one-particle irreducible (1PI) Green functions can be related to each other by taking functional derivatives of this expression, and then setting the fields to zero. An analogous expression, in terms of $W$ and the sources, can be obtained from \re{dZ1} using \re{field-sources}, which will generate relations between connected correlators. The physical contents of the latter are of course identical to that of the equations for 1PI functions. We discuss the relations obtained in terms of 1PI functions, since they show more directly  how the tree-level relation between different terms in the action (the inverse propagator and the vertex) are generalized at the nonperturbative level.   

As a first example, we take the $\delta^2 / \delta \phi_b \delta \chi_c$ derivative of eq.~\re{dZ2}, and integrate by parts in $\eta$ the term coming from the second line. We get
\begin{eqnarray}
& & \!\!\!\!\!\!\!\! \!\!\!\!\!\!\!\! \!\!\!\!\!\!\!\!
 \, \left. \frac{\delta^2 \Gamma}{\delta \varphi_b \left( \bp , \eta' \right) \delta \chi_c \left( \bq , \eta'' \right) }\right|_{\varphi_a = \chi_b = 0}
   \left( \bp \cdot \bw\, \delta_D \left( \eta - \eta' \right) + 
  \bq \cdot \bw\,\delta_D \left( \eta - \eta'' \right) \right) \nonumber\\
 & & \!\!\!\!\!\!\!\! \!\!\!\!\!\!\!\! \!\!\!\!\!\!\!\!
- \int d^3 \bk  \bk \cdot \bw\;\delta_D \left( \bk \right) \partial_\eta \left( {\rm e}^{-\eta}  \, \left.\frac{\delta^3 \Gamma}{\delta \varphi_b \left( \bp , \eta' \right) \delta \chi_c \left( \bq , \eta'' \right) \delta \varphi_2 \left( \bk , \eta \right)} \, \right)\right|_{\varphi_a = \chi_b = 0}
= 0 \, .
\label{G2G3-full}
  \end{eqnarray}
Multiplying by $e^\eta$ and integrating in $\eta$, we get
\beqra
&&G^{-1}_{cb}(p;\eta'',\etap) \,\bp \cdot \bw\,  \left( \,e^{\etap} - e^{\eta''}\right)  \nonumber\\
&& \;\;\;\;\;\;\;\;\;\;\;\;\;\;\;\;\; + \int d\eta\;d^3 \bk \, \bk \cdot \bw\;\delta_D \left( \bk \right) \Gamma^{\chi\vp\vp}_{cb2}(-\bp-\bk,\bp,\bk;\eta'',\etap,\eta) =0\,,
\label{W1}
\eeqra
where we have used the definition of the inverse (fully nonlinear) propagator \cite{MP07b}
\beqra
&&\!\!\!\!\!\left. \frac{\delta^2 \Gamma}{\delta \varphi_a \left( \bp , \eta' \right) \delta \chi_b \left( \bq , \eta'' \right) }\right|_{\varphi_a = \chi_b = 0} \equiv  \delta_D(\bq+\bp)\,\left(g^{-1}_{ba}(\eta'',\etap) - \Sigma_{ba}(p;\eta'',\etap) \right)\,,\nonumber\\
&&\;\;\;\;\;\;\;\;\;\;\;\;\qquad\qquad\qquad\qquad=\delta_D(\bq+\bp)\, G^{-1}_{ba}(p;\eta'',\etap) \,,
\eeqra
and of the full trilinear $\chi \vp \vp$ vertex,
\beqra
&&\!\!\!\!\!\!\!\!\!\!\!\!\!\!\!\!\!\!\!\!\!\!\!\!\!\!\!\!\!\!\left.\frac{\delta^3 \Gamma}{\delta \chi_a \left( \bk , \eta \right) \delta \varphi_b \left( \bp , \eta' \right)  \delta \varphi_c \left( \bq , \eta'' \right)} \, \right|_{\varphi_a = \chi_b = 0} \equiv \delta_D(\bk+\bq+\bp) \Gamma^{\chi\vp\vp}_{abc}(-\bq-\bp,\bp,\bq;\eta,\etap,\eta'')\,,\nonumber\\
&& \;\;\;\;\;\;\;\;\;\;\;\;\;\;\;\;\; = - 2 \delta_D(\eta-\etap)\delta_D(\eta-\eta'')\,e^\eta \, \gamma_{abc}(\bk,\bp,\bq) + O({\mathrm{1- loop}})\,.
\eeqra
Eq.~\re{W1} is one example of the analogous of Ward identities in quantum field theory. It enforces the constraint on fully renormalized ({\it i.e.} nonlinear) quantities (in this case, the inverse propagator, and the trilinear $\chi\vp\vp$ vertex) coming from the underlying symmetry of the theory, in this case galilean invariance. Using the explicit expression for the linear propagator in eq.~\re{gil} and eq.~\re{limitg} for the linear vertex, one can immediately check the identity \re{W1} at tree-level.

\begin{figure}[ht!]
\centerline{
\includegraphics[width=1.\textwidth,angle=0]{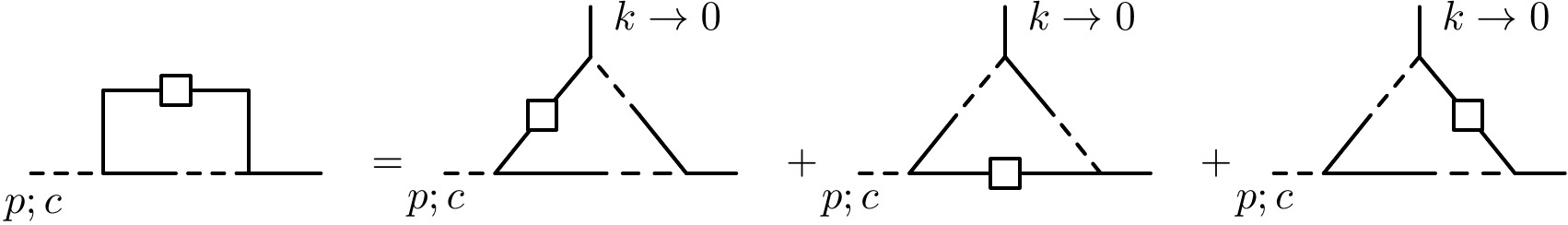}
}
\caption{
Relation  \re{W1} at one loop.
}
\label{fig:w23-1loop}
\end{figure}

At one loop, the relation \re{W1} acquires the diagrammatic form shown in Figure \ref{fig:w23-1loop}, where continuous lines with an open square represent linear PS's and continuous-dotted lines represent propagators, see \cite{MP07b} for details. In the $k\to 0$ limit, and using again the property \re{limitg}, we get that the contributions from the upper tree-level vertex in the three diagrams at the RHS sum up to give

\beq
\!\!\!\!\!\!\!\!\!\
\!\!\!\!\!\!\!\!\!\!\!\!\!\ \frac{\bk\cdot\bl}{k^2}\left(e^{\etap}-1\right)+ \frac{\bk\cdot(\bp+\bl)}{k^2}\left(e^{\eta''}-e^{\etap}\right) -\frac{\bk\cdot\bl}{k^2}\left(e^{\eta''}-1\right) = \frac{\bk\cdot\bp}{k^2}\left(e^{\eta''}-e^{\etap}\right)\,,
\label{sum1l}
\eeq
where $\bl$ is the loop momentum. Since the sum is $\bl$-independent, it factorizes from the loop integral, which gives exactly the 1-loop expression for $\Sigma_{cb}(p;\eta'',\etap)$ and, multiplying eq.~\re{sum1l} by $\bk\cdot \bw \,\delta_D(\bk)$, and integrating in $d^3k$ reproduces the $\bp\cdot\bw \,\left(e^{\eta''}-e^{\etap}\right)$ factor at the LHS of eq.~\re{W1}.

Taking the $\delta^2 / \delta \chi_b \delta \chi_c$  double derivative of \re{dZ2}, and performing similar manipulations as those used in obtaining eq.~\re{W1}, we get another relation between two- and three-point 1PI functions,
\beqra
&&i  \,\Phi_{ab}(q;\eta',\eta'')  \bq \cdot \bw \left( e^{\etap} -e^{\eta''}\right)  \nonumber\\
&& \;\;\;\;\;\;\;\;\;\;\;\;\;\;\;\;\; + \int d\eta\;d^3 \bk \, \bk \cdot \bw\;\delta_D \left( \bk \right) \Gamma^{\chi\chi\vp}_{ab2}(-\bk-\bq,\bq,\bk;\eta'',\etap,\eta) =0\,,
\label{W2}
\eeqra
where \cite{MP07b}
\beq
\!\!\!\!\!\!\!\!\!\!\!\!\!\!\!\!\!\!\!\!\!\!\!\!\!\left. \frac{\delta^2 \Gamma}{\delta \chi_a \left( \bq , \eta \right) \delta \chi_b \left( \bp , \eta' \right) }\right|_{\varphi_a = \chi_b = 0} \equiv  i \delta_D(\bq+\bp)\,\left( P^0(q)u_au_b \delta_D(\eta)\delta_D(\etap) + \Phi_{ab}(q;\eta,\etap)\right)\,,
\eeq
and
\beqra
&&\!\!\!\!\!\!\!\!\!\!\!\!\!\!\!\!\!\!\!\!\!\!\!\!\!\!\!\!\!\!\left.\frac{\delta^3 \Gamma}{\delta \chi_a \left( \bk , \eta \right) \delta \chi_b \left( \bq , \eta' \right)  \delta \varphi_c \left( \bp , \eta'' \right)} \, \right|_{\varphi_a = \chi_b = 0} \equiv \delta_D(\bk+\bq+\bp) \Gamma^{\chi\chi\vp}_{abc}(-\bq-\bp,\bq,\bp;\eta,\etap,\eta'').
\nonumber\\
&&
\eeqra
Notice that the trilinear vertex $\Gamma^{\chi\chi\vp}_{abc}$ is absent at tree-level, but, as 
\beqra
&&\!\!\!\!\!\!\!\!\!\!\!\!\!\!\!\!\!\!\!\!\!\!\!\!\!\!\!\!\!\!\left.\frac{\delta^3 \Gamma}{\delta \chi_a \left( \bk , \eta \right) \delta \chi_b \left( \bq , \eta' \right)  \delta \chi_c \left( \bp , \eta'' \right)} \, \right|_{\varphi_a = \chi_b = 0} \equiv \delta_D(\bk+\bq+\bp) \Gamma^{\chi\chi\chi}_{abc}(-\bq-\bp,\bq,\bp;\eta,\etap,\eta'')\,,
\nonumber\\
&&
\eeqra
it is generated at 1-loop and higher orders.

Ward identities  \re{W1} and \re{W2} can be combined to derive a relation between the nonlinear bispectrum and the nonlinear power spectrum. Indeed, the fully nonlinear bispectrum
\beq
\langle \vp_a(\bk;\eta) \vp_b(\bq;\etap)\vp_c(\bp;\eta'') \rangle \equiv \delta_D(\bk+\bq+\bp) B_{abc}( k, q, p;\eta,\etap,\eta'')\,,
\eeq can be exactly expressed in terms of the full trilinear 1PI vertices, the full PS and the full propagator, as follows 
\beqra
&&\!\!\!\!\!\!\!\!\!\!\!\!\!\!\!\!\!\!\!\!\! B_{abc}(k,q,p;\eta,\etap,\eta'') =\int ds\,ds'\,ds'' \nonumber\\
&&\!\!\!\!\!\!\!\!\!\!\!\!\!\!  \bigg\{- \left[G_{ad}(k;\eta,s)P_{be}(q;\etap,s')P_{cf}(p;\eta'',s'') \Gamma^{\chi\vp\vp}_{def}(\bk,\bq,\bp;s,s',s'') + \mathrm{``2\;cyclic''}\right]\nonumber\\
&&\!\!\!\!\!\!\!\!\!\!\!\!\!\!  - i \left[G_{ad}(k;\eta,s)G_{be}(q;\etap,s')P_{cf}(p;\eta'',s'') \Gamma^{\chi\chi\vp}_{def}(\bk,\bq,\bp;s,s',s'' + \mathrm{``2\;cyclic''}\right]
\nonumber\\
&&\!\!\!\!\!\!\!\!\!\!\!\!\!\!  +  G_{ad}(k;\eta,s)G_{be}(q;\etap,s')G_{cf}(p;\eta'',s'') \Gamma^{\chi\chi\chi}_{def}(\bk,\bq,\bp;s,s',s'')\bigg\} \,,
\label{Bgamma}
\eeqra
where ``cyclic'' refers to summation on the quantities obtained by  permutations of $(\bk,\bq,\bp)$ and, correspondingly, of $(\eta,\etap,\eta'')$ and $(s, s', s'')$. 

Since we are interested in the limit in which one of the external momenta vanishes, we notice that, due to the property of the trilinear vertex coming from translational invariance, $\lim_{k\to 0} \gamma_{abc}(\bk,\bq,\bp)= 0$, the only non vanishing contributions come from terms in which the vanishing momentum is associated to a $\vp$-end of the trilinear vertices. This implies that the third line does not contribute, and that only terms in which the vanishing momentum is carried by a  PS, and not by a  propagator, have to be taken into account.

Multiplying both sides by $\bk \cdot \bw\,\delta_D(\bk)$, and using   the Ward identities \re{W1} and \re{W2}, we get 
\begin{eqnarray}
&&\!\!\!\!\!\!\!\!\!\!\!\!\!\!\!\!\!\!\!\!\! 
 \int d^3 k \bk \cdot \bw \delta_D \left( \bk \right) B_{abc} \left( k , q , \vert  \bq + \bk \vert ; \eta , \eta' \eta'' \right) = 
  \int d s' d s'' u_a P^{(0)} \left( k \right) \bp \cdot \bw  \times \nonumber\\
&& \Bigg[ 
 G_{be} \left( q , \eta' , s' \right) P_{cf} \left( p , \eta'' , s'' \right) 
G^{-1}_{ef} \left( p , s' , s'' \right)  \left(  e^{s''}  - e^{s'} \right) \nonumber\\
&& -  P_{be} \left( q , \eta' , s' \right) G_{cf} \left( p , \eta'' , s'' \right) 
G^{-1}_{fe} \left( q , s'' , s' \right) \ \left( e^{s'} - e^{s''} \right) \nonumber\\
&& -   G_{be} \left( q , \eta' , s' \right) G_{cf} \left( p , \eta'' , s'' \right) 
\Phi_{ef} \left( p , s'' , s' \right)  \left( e^{s''} - e^{s'} \right) \Bigg] \,\,,
\label{wardB-par}
\end{eqnarray}
where we have used the fact that the PS carrying momentum $k$ is in the linear regime. Only the first term in the first line and the first term in the second  line of the expression within square brackets contribute to the final result:
\beq
\!\!\!\!\!\!\!\!\!\!\!\!\!\!\!\!\!\!\!\!\!  \!\!\!\!\!\!\!\!\!\! \int d^3k \,\bk \cdot \bw\,\delta_D(\bk) B_{abc}(k,q,\vert \bq+\bk \vert;\eta,\etap,\eta'') = -\bq \cdot \bw \left(e^{\etap}-e^{\eta''}\right) \, P^0(k) u_a P_{bc}(q;\etap,\eta'')\,,
\label{relbisp}
\eeq
The remaining terms in \re{wardB-par} cancel against each other, as can be seen from the relations \cite{MP07b}
\beqra
&&\int ds\, G_{ac}(q;\eta,s)G^{-1}_{cb}(q;s,\eta') = \delta_{ab} \delta_D(\eta-\etap)\,,\nonumber\\
&& P_{ab}(q;\eta,\etap)=  G_{ac}(q;\eta,0)G_{bd}(q;\etap,0) u_c u_d \,P^0(q) \nonumber\\
&&\;\;\;\;\;\;\;\;\;\;\;\;\;\;\;\;\;\;+\int ds\,ds' \,G_{ac}(q;\eta,s)G_{bd}(q;\etap,s')\Phi_{cd}(q;s,s')\,.
\label{fullPS}
\eeqra
The identity \re{relbisp} can also be expressed as
\beq
\!\!\!\!\!\!\!\!\!\!\!\!\!\ \!\!\!\!\!\!\!\!\!\!\!\!\! \!\!\!\!\!\!\!  \!\!\!\!\! \lim_{k\to0} B_{abc}(k,q,\vert \bq +\bk \vert;\eta,\etap,\eta'') = -  P^0(k) u_a P_{bc}(q;\etap,\eta'') \frac{\bq \cdot \bk}{k^2} \left(e^{\etap}-e^{\eta''}\right)    + O(k^0) \,,
\label{idbisp}
\eeq
where by $O(k^0)$ we indicate terms which do not have the $ q/k$ enhancement and which, in general, do not vanish for $\eta'=\eta''$ and are not proportional to the nonlinear PS $P_{bc}(q;\etap,\eta'')$.

In terms of the density contrast $\delta$,   and going back to conformal time,    this relation rewrites
 \beqra
&&
\lim_{k\to0} B_\delta(k,q,\vert \bq+\bk \vert;\tau,\tau',\tau'') \nonumber\\ 
&&= -  P_\delta^0 (k; \tau, \tau)  P_\delta (q;\tau',\tau'') \, \frac{\bq \cdot \bk}{k^2} 
\frac{D_+ \left( \tau' \right) -  D_+ \left( \tau'' \right) }{D_+ \left( \tau \right)}
    + O(k^0) \,,
\label{idbisp-delta}
\eeqra
where we recall that $D_+ \left( \tau \right)$ is the growth factor at the time $\tau$. This relation is obtained by setting $a=b=c=1$ 
in \re{idbisp} and using eq. \re{doublet}.    Setting some of these indices to $2$ results in  relations between correlators that include also   the velocity field. 

We stress that this  result is exact and nonperturbative, and merely descends from the requirement that the dynamics of the DM fluid  is  galilean invariant. In particular, it holds even in presence of vorticity  and multi streaming, that is, when velocity dispersions and higher order moments cannot be neglected. Indeed, the derivation of the WI and of eq.~\re{idbisp} only requires that the action $S$ and the initial PS in \re{Z} are galilean invariant. As long as these conditions are satisfied, one can also introduce new degrees of freedom, such as velocity dispersion and vorticity, along the lines discussed in \cite{Pietroni:2011iz},  and integrate them over in the path integral without providing a source term for them.  Therefore the range of scales to which this relation can be applied is in principle very large, being limited at very small scales only by the emergence of baryonic physics.

\section{Initial nongaussianity}
\label{fNL}

The relation (\ref{idbisp}) assumes that the initial fields are gaussian. It is interesting to study how it is modified in presence of an initial nongaussianity. For definiteness, we specify the current discussion to  a nonvanishing initial bispectrum. Specifically, adding  in the  generating functional \re{Z} the new term
\begin{eqnarray}
& & \!\!\!\!\!\!\! \!\!\!\!\!\!\! \!\!\!\!\!\!\! \!\!\!\!\!\!\! \!\!\!\!\!\!\! 
Z \left[ J_a ,\, K_b ; \bV0 \right]  = \nonumber\\
&& \!\!\!\!\!\!\! \!\!\!\!\!\!\! \!\!\!\!\!\!\! \!\!\!\!\!\!\!  \!\!\!\!\!\!\!
  \int {\cal D} \varphi_a {\cal D} \chi_b \, {\rm exp } \, \Bigg\{ 
\frac{i}{3!}  \int d^3 \bk   d^3 \bq B_{abc}^0  \left( k , q , \vert \bq +\bk \vert \right) \,
\chi_a \left(  \bk , 0 \right) \chi_b \left(  \bq , 0 \right) \chi_c \left(  -\bk-\bq , 0 \right)
+ \dots\Bigg\} \, ,  \nonumber\\
\label{Z-fNL}
\end{eqnarray}
where dots denotes all the terms present in \re{Z}, we introduce a non vanishing $\chi^3$ interaction in the free action,
\beq
\!\!\!\!\!\!\!  \!\!\!\!\!\!\! \Gamma^{\chi\chi\chi, {\rm tree}}_{abc}(k,q,|\bq+\bk|;\eta,\etap,\eta'') = \delta_D(\eta)\, \delta_D(\etap)\, \delta_D(\eta'') B_{abc}^0  \left( k , q , \vert \bq +\bk \vert \right)\,,
\label{newgamma}
\eeq
which, once attached to external propagators, communicates the effect of the primordial nongaussianity to the full PS, bispectrum, and so on. In particular, at zero-th order in $\gamma_{abc}$, we get the linearly evolved bispectrum of primordial origin,
\beq
\!\!\!\!\!\!\!  \!\!\!\!\!\!\!  B_{abc}^{\rm tree} \left( k , q , \vert \bq+\bk \vert ; \eta , \eta' , \eta'' \right) = g_{ad}(\eta)g_{be}(\etap)g_{cf}(\eta'') \,B_{cde}^0  \left( k , q , \vert \bq +\bk \vert \right)\,.
\label{treeB}
\eeq

In the following, we assume that, in the squeezed limit, the initial bispectrum factorizes a term proportional to the initial power spectrum of the non vanishing momentum,
\begin{equation}
\lim_{k\to 0} B_{abc}^0  \left( k , q , \vert \bq +\bk \vert \right) = f_a \left( k \right) P^0_{bc} \left( q ; 0 , 0 \right) \,.
\label{B0-P0}
\end{equation}
The initial nongaussianity is typically given in terms of a gauge invariant variable in the super horizon regime. For example, in terms of  Bardeen's gauge invariant primordial gravitational potential we have
\begin{equation}
\lim_{k\to0} B_\Phi \left( k , q , \vert \bq+\bk \vert \right) = 4 f_{\rm NL} P_\Phi \left( k \right) P_\Phi \left( q \right)\,,
\label{local-Phi}
\end{equation}
where $ f_{\rm NL} =  f_{\rm NL}^{\rm local}$ if the initial nongaussianity is precisely of the local form, while a different numerical factor is present for different other initial shapes.  This relation is verified if the initial conditions are obtained from models of single field inflation \cite{Maldacena:2002vr}.

In the linear regime, Bardeen's potential is related to our field $\vp_a$ by the following relation, 
\begin{equation}
\vp_a^{\rm linear}  \left( k , \eta \right) = \frac{2 k^2 T \left( k \right)}{3 \Omega_{m,0} H_0^2 {\rm e}^{\eta_0}} \, \Phi \left( k \right)\,,
\end{equation}
where the suffix $0$ at the denominator refers to the current time (where $H_0 = {\cal H}_0$, given that the scale factor is normalized to $1$ today), and where $T \left( k \right)$ is the transfer function of matter fluctuations, normalized such that $T \left( k \right) \rightarrow 1$ when $k \rightarrow 0$. Therefore, the relation (\ref{treeB}) can be rewritten as a relation between the tree-level bispectrum and the linear PS's,
\begin{equation}
 \!\!\!\!\!\!\!  \!\!\!\!\!\!\! \!\!\!\!\!\!\!  \!\!\!\!\!\!\!
\lim_{k\to0} B_{abc}^{\rm tree} \left( k , q , \vert \bq+\bk \vert ; \eta , \eta' , \eta'' \right) =  \frac{6 f_{\rm NL} 
 \Omega_{m,0} H_0^2 {\rm e}^{\eta_0} }{k^2 T \left( k \right) } P_{aa}^0 \left( k; \eta, \eta \right)    P_{bc}^0 \left( q; \eta' , \eta'' \right)\,,   
 \label{local-Phi}
\end{equation}
which is of the form  \re{B0-P0} with 
\beq f_a \left( k \right) =  \frac{6 f_{\rm NL} 
 \Omega_{m,0} H_0^2 {\rm e}^{\eta_0} }{k^2 T \left( k \right) } 
P^0 \left( k \right) u_a     \,,
\label{fa}
\eeq
 (notice that $u_a = 1$).

We are interested in how the relation  \re{idbisp} between the fully nonlinear bispectrum and PS's is modified at first order in the initial nongaussianity (given that the initial nongaussianity, if present, is small, higher order  corrections will be very subdominant). Therefore, at the full nonlinear level in $\gamma_{abc}$, we should consider all the diagrams that contribute to the bispectrum that have at most one  new vertex of the type given in eq.~\re{newgamma}. The additional term in \re{Z-fNL}, involving equal-time fields,  is Galilean invariant, and so it does not modify the Ward identities \re{W1} and \re{W2}. Also the formal relations \re{Bgamma} and \re{fullPS} are unmodified. The additional vertex however modifies the quantities $P_{ab}, G_{ab}, \Phi_{ab}, \Gamma_{abc}$ that enter in these relations.

\begin{figure}[ht!]
\centerline{
\includegraphics[width=1.\textwidth,angle=0]{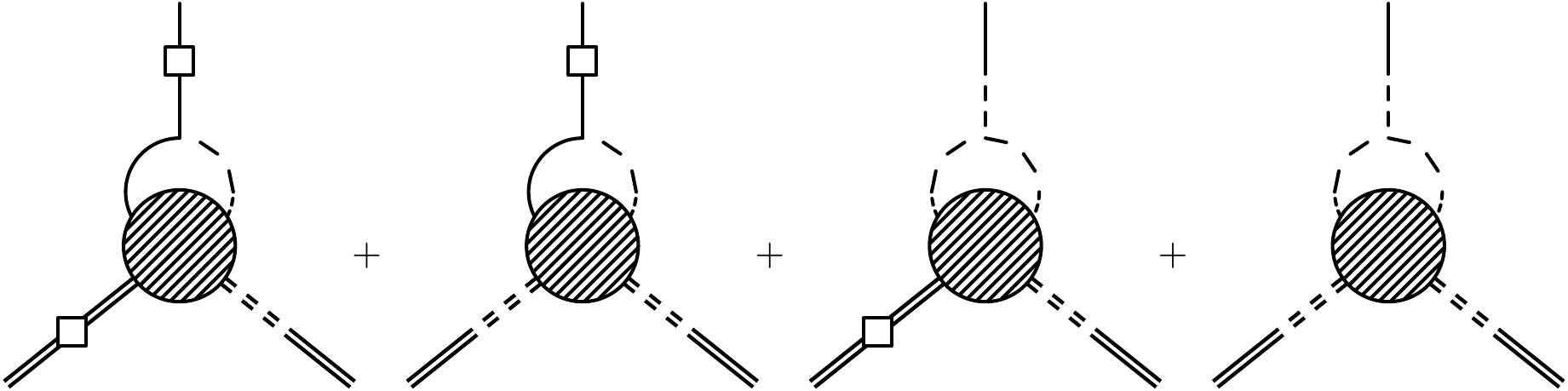}
}
\caption{
Diagrams contributing to the nonlinear  bispectrum in the squeezed limit, in which the top leg carries vanishing momentum.  Only terms up to linear order in the initial nongaussianity are included. The external double lines denote the full renormalized PS and propagator, while the field in the upper vertex is still in the linear regime.
}
\label{fig:fNL}
\end{figure}

We discuss separately the two possible cases: either the three $\chi$ ends of the additional interaction \re{newgamma} are all linked to internal vertices in the loops, or one of them is directly connected to the external $\vp_a \left(  \bk \right)$ through a propagator (where $\bk$ is the smallest momentum in the bispectrum). In the first case, the contributions surviving the $k \to 0$ limit correspond to the first two diagrams in Fig.~\ref{fig:fNL}, which we denote by $B^I$. The second case corresponds to the last two diagrams in the figure, which we denote by $B^{II}$. The full bispectrum in the squeezed limit is then given by $B^I + B^{II}$ up to second and higher order corrections in the initial nongaussianity.
  
Let us first discuss the $B^I$ contribution. The first and second diagrams in   Fig.~\ref{fig:fNL} correspond, respectively,  to those terms in the second and third lines of eq.~\re{Bgamma} in which the vanishing $k$ momentum is carried by an external PS, not by a propagator. As in the previous section, $B^I$ does not receive contributions from the fourth line of  eq.~\re{Bgamma} and therefore it will be related to the nonlinear PS  by the same identity \re{idbisp},
\beq
\!\!\!\!\!\!\!\!\!\!\!\!\!\ \!\!\!\!\!\!\!\!\!\!\!\!\!\ \lim_{k\to0} B_{abc}^I (k,q,\vert \bq + \bk \vert ;\eta,\etap,\eta'') = -  P^0(k) u_a P_{bc}(q;\etap,\eta'') \frac{\bq \cdot \bk}{k^2} \left(e^{\etap}-e^{\eta''}\right)     \,.
\label{BI}
\eeq
 The difference between this expression and \re{idbisp} is that now both the bispectrum and the full PS include also
 linear contributions in the initial nongaussianity.

 Let us now discuss the  $B^{II}$ part.  A generic diagram that contributes to it has the amplitude
 \begin{eqnarray}
&&  \!\!\!\!\!\!\!\!\!\!\!\!\!\ \!\!\!\!\!\!\!\!\!\!\!\!\!\  \!\!\!\!\!\!\!\!\!\!\!\!\!\  \!\!\!\!\!\!\!\!\!\!\!\!\!\ 
\lim_{k\to0}  \left\langle \vp_a \left( \bk , \eta \right) \frac{i}{3!} B^0_{def} \left( k', q', \vert -\bq'-\bk' \vert \right) \chi_{d} \left( \bk', 0 \right) 
  \chi_{e} \left( \bq', 0 \right)  \chi_{f} \left( - \bk' - \bq' , 0 \right) \dots  \vp_b \left( \bq , \eta' \right)  \vp_c \left( - \bk - \bq , \eta' \right) 
\right\rangle \nonumber\\
&&  \!\!\!\!\!\!\!\!\!\!\!\!\!\ \!\!\!\!\!\!\!\!\!\!\!\!\!\  \!\!\!\!\!\!\!\!\!\!\!\!\!\   \!\!\!\!\!\!\!\!\!\!\!\!\!\ 
= \left[ \lim_{k\to0}  
 g_{ad} \left( \eta \right) 
  \frac{6 f_{\rm NL} 
 \Omega_{m,0} H_0^2 {\rm e}^{\eta_0} }{k^2 T \left( k \right) } P^0 \left( k   \right)  u_{d}
  \right] \left\langle \frac{-1}{2} P^0_{ef} \left(  q' , 0 , 0 \right)
  \chi_{e} \left( \bq', 0 \right)  \chi_{f} \left(   - \bq' , 0 \right) \dots  \vp_b \left( \bq , \eta' \right)  \vp_c \left(   \bq , \eta' \right) 
\right\rangle \,. \nonumber\\
\label{B2full-Pfull}
\end{eqnarray} 
 In the first line, only the external fields and the $B^0 \chi^3$ vertex have been written explicitly; to obtain the second line we have contracted $\vp_a \left( \bk \right)$ with one of the $\chi$ from this interaction with a linear propagator,  we have used  (\ref{B0-P0}), and we have disregarded $\bk$ when summed with $\bq$. The second expression is the amplitude of a diagram contributing to the full power spectrum $P_{bc} \left( q ; \eta' \eta'' \right)$ of the short modes at ${\rm O } \left( f_{\rm NL}^0 \right)$ times a function ($ \lim_{k\to0}   g_{aa'} \left( \eta \right)  f_{a'} \left( k \right)  $) that depends only on the time and momentum of the soft  mode $\vp_a \left( \bk , \eta \right)$.

\begin{figure}[ht!]
\centerline{
\includegraphics[width=0.8\textwidth,angle=0]{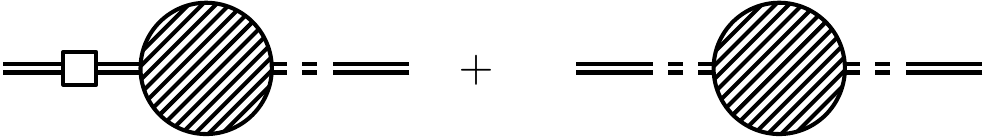}
}
\caption{
Diagrams contibuting to the full PS.
}
\label{fig:PS}
\end{figure}

An alternative way to see this is to realize that, in the squeezed limit of eq.~\re{B0-P0}, the new term of eq.~\re{Z-fNL} can be thought as a shift of the linear PS in the gaussian case of eq.~\re{Z}  to 
\beq
P^0_{bc}(q) \to P^0_{bc}(q) \left(1 - i f_a(k) \chi_a(\bk,0)\right)\,,
\label{shift}
\eeq
with $k$ the vanishing mode. Therefore, all diagrams that contribute to $B^{II}$ (at $O(f_{NL})$) can obtained by starting from a given diagram contributing to the full PS at ${\rm O } \left( f_{\rm NL}^0 \right)$, and by replacing one of the linear PS that enter in this diagram by the second term at the RHS of eq.~\re{shift}. The remaining $ \chi_a(\bk,0)$ field is then connected to the external soft $\vp_a$ field by a linear propagator.

Therefore we get the nonlinear relation 
\begin{equation}
\lim_{k \to 0} B_{abc}^{II} \left( k , q , \vert \bq + \bk \vert ; \eta , \eta' , \eta''  \right) = f_a(k)    \,  P_{bc} \left( q ; \eta' , \eta'' \right)   \,.
\label{BII}
\end{equation}
We stress that this is a relation between the full contribution to $B^{II}$ in the squeezed limit  and the full power spectrum of the  short modes. 

Once written in terms of the density contrast $\delta$, the two relations \re{BI} and \re{BII} give the full bispectrum
\begin{eqnarray}
&&  \!\!\!\!\!\!\!\!\!\!\!\!\!\ \!\!\!\!\!\!\!\!\!\!\!\!\!\  \!\!\!\!\!\!\!\!\!\!\!\!\!\ 
\lim_{k \to 0} B_\delta \left( k , q , \vert \bq + \bk \vert ; \tau , \tau' , \tau'' \right) \nonumber\\
&&  \!\!\!\!\!\!\!\!\!\!\!\!\!\ \!\!\!\!\!\!\!\!\!\!\!\!\!\  \!\!\!\!\!\!\!\!\!\!\!\!\!\ 
= \left[ -  \, \frac{\bq \cdot \bk}{k^2} 
\frac{D_+ \left( \tau' \right) -  D_+ \left( \tau'' \right) }{D_+ \left( \tau \right)}
+  \frac{6 f_{\rm NL}  \Omega_{m,0} H_0^2  }{k^2 T \left( k \right) } 
 \frac{D_+ \left( \tau_{\rm 0} \right)  }{D_+ \left( \tau \right)}
\right] P_\delta (k; \tau, \tau)  P_\delta (q;\tau',\tau'') 
  + {\rm O } \left( k^0 ,\; f_{\rm NL}^2 \right)  \,, \nonumber\\
 \label{P-B-finalid}
\end{eqnarray}
in terms of the full  PS of the short modes, (where we have used the explicit expression for $f_a(k)$, eq.~\re{fa}, and the fact that $u_a = 1$).

\section{Galieian invariance and perturbation theory}
\label{PTG}

Any computational scheme should reproduce the transformation properties of the correlators under a GT, eq.~\re{general-corr-GT}, in particular giving frame independent results for  equal-time correlators. Since a GT can be seen as a coherent motion on very large (actually, infinite) scales, a non GT-invariant scheme would generally introduce a spurious dependence of small scale modes at momentum $q$ on very large scale modes of momentum $k\ll q$, typically enhanced by powers of $q/k$ over the true physical dependence. The role of the Ward identities is actually to guarantee that no such spurious effect are present in  a given approximation scheme. 

In Sect.~\ref{Ward} we have seen how standard PT  (up to 1-loop order) fulfills the Ward identities. In this section we want to investigate how the GT for correlators are obtained in standard PT and in particular, how the invariance is achieved for equal-time correlators.
Then, in Sect.~\ref{resummations} we will discuss the same issues for some of the resummation methods proposed so far.

As we have seen in Sect.~\ref{PIn}, in a frame with non-zero average velocity, $\bV0$, the field $\vp_a$ has a non-vanishing expectation value --  a {\it tadpole} in field theoretical language -- given by $\bar\vp_a$ in eq.~\re{phi2-bar}.  This adds a new Feynman rule to the set of rules for the c.o.m. frame ($\bV0=0$) which, when properly taken into account, reproduces the transformation law of eq.~\re{general-corr-GT}.

\begin{figure}[ht!]
\centerline{
\includegraphics[width=0.5\textwidth,angle=0]{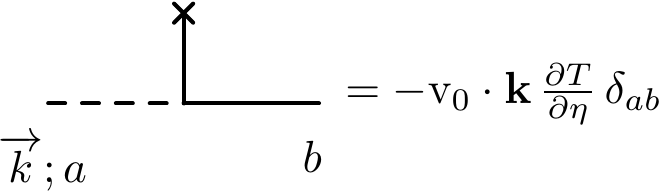}
}
\caption{
Inserting the tadpole in all  possible ways in a diagram,  integrating over the corresponding times, and summing over all diagrams contributing to a correlator, gives the GT of that correlator.
}
\label{fig:GT-massins}
\end{figure}

A convenient way to see this, is to rotate away the phases from the sources at the RHS of eq.~\re{id1}, by performing the change of integration variables $\vp_a(\bk,\eta)\to \vp_a(\bk,\eta) e^{-i \bk \cdot \bV0 T}$, $\chi_a(\bk,\eta)\to \chi_a(\bk,\eta)e^{-i \bk \cdot \bV0 T}$. It gives
\beq
\!\!\!\!\!\!\!\! \!\!\!\!\!\!\!\! \!\!\!\!\!\!\!\! \!\!\!\!\!\!\!\!  \!\!\!\!
Z \left[ J_a,  K_b; \bV0 \right] = 
{\rm e}^{i \int d \eta d^3 \bk J_2 \left( - \bk , \eta \right) {\bar \varphi}_2 \left( \bk, \eta , \bV0 \right)}{\rm e}^{\int d \eta d^3 \bk \frac{\delta^2}{\delta K_a \left( \bk ,\, \eta \right) \delta J_a \left( - \bk ,\, \eta \right) } \bk \cdot \bV0 
\partial_\eta T \left( \eta \right) }  
Z \left[ J_a,  K_b;{\bf 0 } \right] \, .
\label{id2}
\eeq
The effect of the second exponential at the RHS can be seen by expanding in perturbation theory in the interaction vertex $\gamma_{abc}$ (the first exponential does not contribute to correlators with non vanishing external momenta),
by using
\beq
Z \left[ J_a,  K_b;{\bf 0 } \right] = {\rm e}^{ \int d \eta d^3 \bk d^3 \bq d^3 \bp \,\gamma_{abc}(\bk,\bq,\bp)\frac{\delta^3\;\;\;}{\delta K_a(\bk,\eta)\delta J_b(\bq,\eta)\delta J_c(\bp,\eta)}} Z_0 \left[ J_a,  K_b;{\bf 0 } \right] \,,
\label{pertexp}
\eeq
where the free functional is obtained by setting $\gamma_{abc}=0$ in the path integral and performing the gaussian integral over $\vp_a$ and $\chi_a$, \cite{MP07b}
\begin{eqnarray}
& & 
\!\!\!\!\!\!\!\!\!\!\!\!\!\!\!\! \!\!\!\!\!\!\!\!\!\!\!\!\!\!\!\! 
Z_0[J_a,\, K_b;{\bf 0}] =  {\rm exp} \Bigg\{ - \int d^3 \bk d \eta d \etap \Bigg[ \frac{1}{2} J_a \left( \bk , \eta \right) P_{ab}^0 \left( k ; \eta , \etap \right) J_b \left( - \bf k , \etap \right)  \nonumber\\
&& \quad\quad\quad\quad\quad\quad\quad\quad
  + i J_a \left( \bk , \eta \right) g_{ab} \left( k ; \eta , \etap \right) K_b \left( - \bk , \etap \right) 
 \Bigg]   \Bigg\} \, . 
      \label{z0free}
\end{eqnarray}
The diagrammatic meaning of eqs.~\re{id2} and \re{pertexp} is straightforward: starting from a given diagram for a given correlator in the $\bV0= 0$ frame and then inserting in all possible ways the tadpole given in Fig.~\ref{fig:GT-massins} (weighting the insertion of $n$ tadpoles by $1/n!$),    
the corresponding diagram for the  $\bV0 \neq 0$ frame is obtained.

\begin{figure}[ht!]
\centerline{
\includegraphics[width=0.4\textwidth,angle=0]{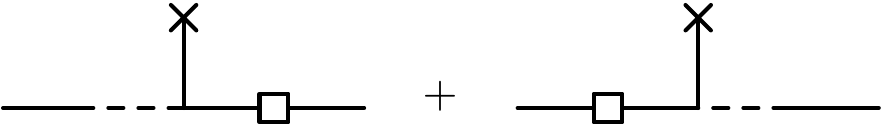}
}
\caption{
One tadpole insertions on a linear power spectrum reproduce the first  order term in the expansion in $\bV0$ of the linear part    of \re{power-GT}. 
}
\label{fig:1mass-PS}
\end{figure}

Let us show how it works explicitly by considering  the GT for the two-point correlator ({\it i.e.} the PS between unequal times), which, as we have shown in eq. \re{general-corr-GT}, transforms as
\begin{equation}
P_{ab} \left( q; \eta, \etap  \right) \; \rightarrow \; {\rm e}^{-i \bq \cdot \bV0 \left[ T \left( \eta \right) - T \left( \etap \right) \right] }  \, P_{a b} \left( q; \eta , \etap \right) \, .
\label{power-GT}
\end{equation}
Let us first discuss the transformation of the linear PS, given in the $\bV0=0$ frame by
\beq
P_{ab}^0 \left( q; \eta, \etap  \right) = g_{ac}(\eta)g_{bd}(\etap)P^0_{cd}(q;0,0) = P^0(q) u_a u_b\,,
\eeq
where we have used eq.~\re{Plin} and the property of the linear propagator, $g_{ab}(\eta)u_b = u_a \theta(\eta)$.  Inserting the tadpole in the two possible ways, namely on the left and on the right legs, as in Fig. \ref{fig:1mass-PS} we get
\beqra
&&  -  i\,\bq \cdot \bV0
\int d s \partial_s T(s) \left[
g_{a c} \left(    \eta - s \right) P_{c b}^0 \left( q;s , \etap \right) - 
 P_{a c}^0 \left(q, \eta , s \right)  g_{b c} \left(    \etap-s \right) \right]\nonumber\\
 && =  -  i\,\bq \cdot \bV0  \left(T(\eta)-T(\etap)\right) P^0(q)u_a u_b\,,
\eeqra
which reproduces the contribution to the transformation in eq.~\re{power-GT} at the linear order in $\bV0$. 

\begin{figure}[ht!]
\centerline{
\includegraphics[width=0.7\textwidth,angle=0]{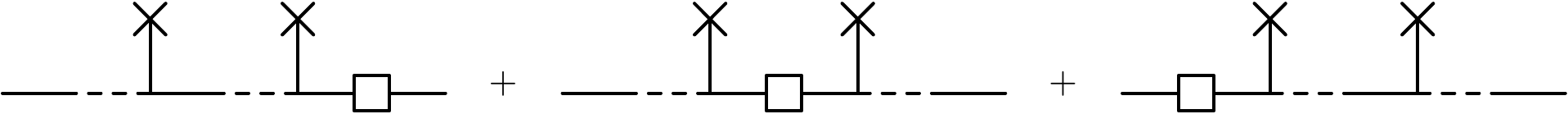}
}
\caption{
Two tadpole insertions on a linear power spectrum reproduce the second  order term in the expansion in $\bV0$ of the linear part    of \re{power-GT}. 
}
\label{fig:2mass-PS}
\end{figure}

The $O(\bV0^2)$ term is given by the three diagrams in Fig.~\ref{fig:2mass-PS}, which give respectively the following three lines 
\begin{eqnarray}
& & \!\!\!\!\!\!\!\! \!\!\!\!\!\!\!\!\!\!       
 \left( -i \, \bq \cdot \bV0 \right)^2 \int d s \partial_s T(s)  ds'  \partial_{s'} T(s') 
\Bigg\{  g_{a c} \left( \eta-s\right)  g_{cd} \left( s-s'\right) 
  P_{d b}^0 \left(q; s' , \etap \right)
    \nonumber\\
& & \qquad\qquad\;\;\;\;\;\;\;\quad\quad  \quad\quad   \quad\quad    -  g_{a c} \left(\eta-s \right) P_{cd}^0 \left(q;s,s' \right) 
  g_{b d} \left( \etap-s' \right)  \nonumber\\
& &\qquad\qquad\;\;\;\;\;\;\;\quad\quad  \quad\quad   \quad\quad 
  + P_{a c}^0 \left(q; \eta , s \right)  g_{dc} \left(s'-s\right) 
g_{bd} \left( \etap-s' \right) \Bigg\}     \, .
 \nonumber\\
& &  \!\!\!\!\!\!\!\! \!\!\!\!\!\!\!\! \!\!\!\!\!\!\!\!  
\label{P-GT-01}
\end{eqnarray}

Using the composition relation between the linear propagators
\begin{equation}
g_{ab} \left( \eta-s \right) \, g_{bc} \left( s-\etap \right) = g_{a c} \left( \eta-\etap \right)\theta \left( \eta-s \right) \theta \left( s- \etap \right) \, , 
\label{composition}
\end{equation}
eq.~ \re{P-GT-01} rewrites 
\begin{eqnarray}
& &   \!\!\!\!\!\!\!\! \!\!\!\!\!\!\!\!  \!\!\!\!\!\!\!\! \!\!\!\!\!\!\!\! \!\!\! \!\!\!
\left( -i \, \bq \cdot \bV0 \right)^2  \left[
 \int_0^{\eta}   d s
   \int_0^{s}   d s'  -  
 \int_0^{\eta}   d s
 \int_0^{\etap}   d s' +   
 \int_0^{\etap}   d s'     
 \int_0^{s'}   d s \right]\partial_s T(s)\,\partial_{s'}T(s')\, P^0(q)u_a u_b
\nonumber\\
& & 
  = \frac{1}{2} \left[ -i \, \bq \cdot \bV0 (T(\eta)-T(\etap)) \right]^2 P^0(q)u_a u_b\,.
\label{PS012}
\end{eqnarray}
The higher orders in $\bV0$ are reproduced by iterating the same procedure and using the property of the time integrals
\beq
\!\!\!\!\!\! \!\!\!\!\!\!\!\!\!\!\!\!   \int_0^\eta ds_1 \partial_{s_1} T(s_1) \int_0^{s_1} ds_2 \partial_{s_2} T(s_2) \dots \int_0^{s_{m-1}} ds_m \partial_{s_m} T(s_m) = \frac{\left[T(\eta)-T(0)\right]^m}{m!}\,.
\eeq

The derivation above can be generalized beyond tree-level in perturbation theory. The crucial point is to realize how the tadpole insertions behave for closed loops. Indeed, by the same mechanism already seen in eq.~\re{sum1l}, inserting in all possible ways the tadpole
 in the internal legs of a closed loop, and integrating over the time of the insertion, one obtains a contribution that is independent on the loop momentum, given by
\beq
-i \,\sum_i \bp_i\cdot \bV0\;T(s_i),
\eeq
where $\bp_i$ and $s_i$ are the incoming momentum and the time argument of the external legs attached to the loop. Using this fact, one can realize that the contributions  from the tadpoles factorize from a given loop diagram, provided they are attached in all possible ways to all the lines, and the transformation in eq.~\re{power-GT} (as that for any other correlator) is reproduced. Notice that this transformation holds for any diagram independently, that is, it is not necessary to sum all the diagrams contributing to the correlator at a given perturbative order to get the transformation in  eq.~\re{power-GT}  right. However, it does not mean that it is consistent, from the GI point of view, to discard some diagrams with respect to others of the same perturbative order, as we will discuss now.

\begin{figure}[ht!]
\centerline{
\includegraphics[width=0.7\textwidth,angle=0]{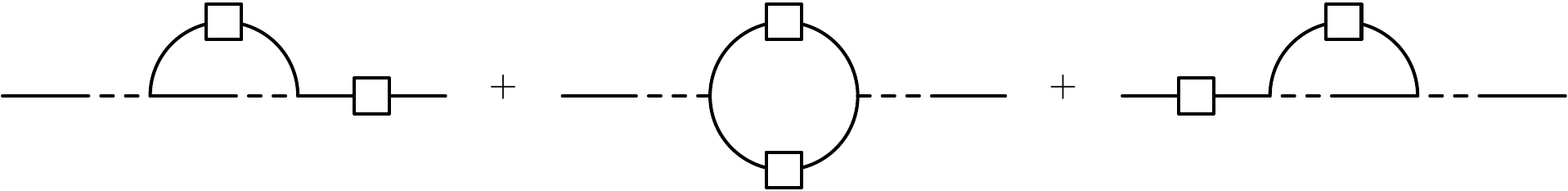}
}
\caption{
One loop contributions to the power spectrum. 
}
\label{fig:2loop-PS}
\end{figure}

Consider again the two-tadpole insertions of Fig.~\ref{fig:2mass-PS}. The three diagrams can be thought as the result of the following operation: take the 1-loop contribution to the PS, given in Fig.~\ref{fig:2loop-PS}, and ``open" one linear PS in all possible ways, namely, replace one linear PS, $\langle\vp_a(\bq;s) \vp_b(\bp;s') \rangle=\delta_D(\bq+\bp)P_{ab}^0(q;s,s')$ with the product $\bar\vp_a(\bq;s) \bar\vp_b(\bp;s')$, where $\bar\vp_a$ is given in eq.~\re{phi2-bar}~\footnote{Notice that, in the intermediate diagram of Fig.~\ref{fig:2loop-PS}, it is possible to open two internal power spectra, obtaining the same result. This compensate the fact that this diagram enters with a relative $1/2$ multiplicity with respect to the other two diagrams in the Figure, while all the diagrams of Figure \ref{fig:2mass-PS} have the same multiplicity.}. This is indeed the effect of a change of frame, since, as we have seen, it corresponds to giving the expectation value $\bar\vp_a$ to $\vp_a$. Put in this way, we realize that the $O(\bV0^2)$  contribution of the GT of the tree-level PS can be obtained by taking the complete set of 1-loop diagrams and opening one linear PS. 

This can be generalized at all loops and at all orders in $\bV0$.
If we start from all possible diagrams with $L$ loops that contribute to a given correlator, and we open $P$ linear power spectra of that diagram in all possible ways,  we end up with all possible diagrams with $L-P$ loops contributing to that correlator, multiplied by the $O(\bV0^{2P})$ term of the Taylor expansion of the GT phase, up to an overall combinatorial factor $(2 P)!!$. This fact provides us a useful way to check the galilean invariance of a given approximation scheme. 

If we consider a GT-invariant correlator, such as the equal-time PS, then the sum of the opened diagrams should cancel. Physically, the cancellation reflects the fact that equal time correlators are insensitive to a global boost of particles velocities (see eq.~\re{general-corr-GT} and next section). Therefore, locking the expansion in $\bV0$ to the PT expansion in $\gamma_{abc}$, as suggested by eqs.~\re{id2} and \re{pertexp}, provides the insightful link between  GI and IR sensitivity, first elucidated in \cite{Scoccimarro:1995if}.
This link results in non-trivial relations between different contributions to a given quantity at a fixed PT order. In the 1-loop PS discussed before, the three diagrams of Fig.~\ref{fig:2loop-PS} exhibit a nontrivial cancellation between the mode-mode coupling diagram in the middle and the other two diagrams, in which an external propagator is renormalized.

If the opening procedure in a given set of diagrams for a galilean invariant correlator does not result in a complete cancellation, then that set is not insensitive to the effects of a GT transformation.

\section{Galielian invariance and IR sensitivity}
\label{GTIR}

Lack of invariance under GT (in the locked $\bV0$-$\gamma_{abc}$ expansion discussed at the end of the previous section)  in a given calculational scheme can result in a spurious dependence of the short scale dynamics on large scales. To see this, we introduce a window function $W \left( \lambda k \right) \simeq \theta \left( 1 - k \lambda \right)$ (namely a regular function that however introduces a sharp difference between short and large wavelength modes), and define
\begin{equation}
\vp_a^L \left( \bk,\eta \right) \equiv W \left( \lambda k \right) \vp_a \left( \bk,\eta \right) \;\;,\;\;
\vp_a^S \left( \bk,\eta \right)= \vp_a\left( \bk,\eta \right) - \vp_a^L\left( \bk,\eta \right) \;,
\label{L/S}
\end{equation}
and identically for the field $\chi_a$ and for the sources $J_a$ and $K_b$. For definiteness, we assume that we are in the frame of vanishing average velocity, $\bV0=0$.

We are interested in the dynamics of the short wavelength modes in the background of the long wavelength modes. Under the  separation \re{L/S}, the   cubic term $- {\rm e}^\eta \gamma_{abc} \chi_a \vp_b \vp_c$ then gives rise to interactions between these two types of modes. The interactions involving   two long and one short  wavelength modes vanish, due to the momentum conservation in the vertex. We obtain
\begin{eqnarray} & &  \!\!\!\!\!\!\!\! \!\!\!\!\!\!\!\! \!\!\!\!\!\!\!\!  \!\!\!\!\!\!\!\!  
i S_{\rm int,LS} = - i \int d \eta {\rm e}^\eta d^3 \bk d^3 \bp d^3 \bq \gamma_{abc} \left( - \bk , - \bp , -\bq \right)\Bigg[ \chi^L_a \left( \bk , \eta \right)  \varphi^L_b \left( \bp , \eta \right)   \varphi^L_c \left( \bq , \eta \right) 
\nonumber\\
& &  \!\!\!\!\!\!\!\! \!\!\!\!\!\!\!\! \!\!\!\!\!\!\!\! \qquad\qquad\qquad+\chi^L_a \left( \bk , \eta \right)  \varphi^S_b \left( \bp , \eta \right)   \varphi^S_c \left( \bq , \eta \right) + 
2  \chi^S_a \left( \bk , \eta \right)  \varphi^L_b \left( \bp , \eta \right)   \varphi^S_c \left( \bq , \eta \right) \Bigg] \, .
\label{Sint-LS}
\end{eqnarray}
The first term involves only long wavelength and approximately linear modes, and can therefore be neglected in this discussion, the second one is suppressed as $k\ll p,q$ (since the vertex vanishes for $k\rightarrow 0$). As for the third term, we assume that the long wavelength mode is still in the linear regime, and therefore we can express it  as
\begin{equation} 
\varphi_b^L \left( \bp , \eta \right) \simeq - {\rm e}^{- \eta} \frac{\theta(\bp)}{{\cal{H}}f} u_b=   - i {\rm e}^{- \eta} \partial_\eta T \,\bp \cdot {\bf V} \left( \bp \right) \, u_b \, ,
\end{equation}
where ${\bf V}  \left( \bp \right)$ is the Fourier transform of the physical velocity (corresponding to a peculiar velocity $\dot T \,{\bf V} (\bp))$. The last term in \re{Sint-LS} is then proportional to
\begin{equation}  \!\!\!\!\!\!\!\! \!\!\!\!\!\!\!\! \!\!\!\!\!\!\!\!   \!\!\!\!\!\!\!\!    
 \gamma_{a1c} \left( - \bk , - \bp , -\bq \right) +  \gamma_{a2c} \left( - \bk , - \bp , -\bq \right)
= \left[  \frac{\bp \cdot \bq}{2 p^2} \delta_{ac} + {\rm O } \left( 1 \right) \right] \delta_D \left( \bk + \bp + \bq \right)
 \;\;,\;\; p \ll q \, .
\label{Taylor-gamma}
\end{equation}
The first term in this expression dominates once inserted into  \re{Sint-LS},  giving 
\begin{eqnarray}
\!\!\!\!\!\!\!\!  i S_{\rm int,LS} &  \simeq &  -  \int_{p \ll q} d \eta \partial_\eta T \,    d^3 \bp d^3 \bq \, \frac{\bp \cdot \bq \, \bp \cdot {\bf V} \left( \bp \right)}{p^2 } \chi^S_a \left( - \bq - \bp  , \eta  \right)     \varphi^S_a \left( \bq , \eta  \right)  + \dots \nonumber\\
\!\!\!\!\!\!\!\! & \simeq & \int d \eta d^3 \bq  \left[  - \partial_\eta T  \bq \cdot \int_{p \ll q} d^3 \bp \, {\bf V} \left( \bp \right) \right]
 \chi^S_a \left( - \bq    , \eta  \right)     \varphi^S_a \left( \bq , \eta  \right)  + \dots \nonumber\\
\label{Sint-LS0}
\end{eqnarray}
where we have written explicitly only the contribution from the first term at the RHS of \re{Taylor-gamma}. However, once inserted in a correlator for the short wavelength modes, this expression has exactly the same effect as the tadpole term at the exponent of eq.~\re{id2}, as it becomes even more clear if we make the replacement ${\bf V(p)} \to\bV0\, \delta_D(\bp)$.
Not surprisingly, from the point of view of short modes, the contribution of a {\em velocity} perturbation on ultra long modes is indistinguishable from that of a GT, and therefore it should have no effect on correlators that are galilean invariant. If GI is not properly taken into account, this spurious dependence of short scales on long scales can easily dominate over the physical one, induced by a long wavelength {\em density} perturbation as discussed, for instance, in \cite{Baldauf:2011bh}, which, being proportional to $\delta_D(\bp)$, and not to  $\bp\, \delta_D(\bp)$ as the velocity one, gives  $O(1)$ terms from eq.~\re{Taylor-gamma}. This would be the case, for instance, if, in the computation of the PS, one would neglect one of the diagrams in Fig.~\ref{fig:2loop-PS} and keep the other two.

\section{A galielian invariant resummation scheme}
\label{resummations}
As we have discussed in Sects.~\ref{PTG} and \ref{GTIR}, standard PT respects GI: the computation of an equal time correlator at a finite PT order gives frame independent results once all the diagrams for that correlator at that order are taken into account. All different resummation methods, like RPT \cite{RPTa}, the multipoint propagator expansion \cite{Bernardeau:2008fa}, closure equations \cite{TaruyaIII}, or the time-evolution equations of \cite{Anselmi:2012cn}, amount to a reorganization of the perturbative expansion: in these schemes, a finite order approximation selects certain classes of PT contributions while neglecting others  of the same PT order. Performing the computation in the new scheme at all orders would of course coincide with computing the result at  infinite order in standard PT, and would therefore respect galilean invariance, but this is not automatically guaranteed when a finite order in the approximation is considered, as is mandatory on practical terms.

Take for instance the computation of the PS in the RPT scheme discussed in  \cite{RPTBAO}. The structure of the PS is the same as the exact expression in the second of eqs.~\re{fullPS}, namely,
\beqra
&& P_{ab}(q;\eta,\etap)=  G_{ac}(q;\eta,0)G_{bd}(q;\etap,0) u_c u_d \,P^0(q) \nonumber\\
&&\;\;\;\;\;\;\;\;\;\;\;\;\;\;\;\;\;\;+\int ds\,ds' \,G_{ac}(q;\eta,s)G_{bd}(q;\etap,s')\Phi_{cd}(q;s,s')\,,
\label{fullPS1}
\eeqra
 in which the full propagator $G_{ab}(k;\eta,s)$ is replaced by an expression interpolating between 1-loop PT at small $k$ and the large-$k$ result \cite{RPTb}, 
\beq
G^{eik}_{ab} \left(k,  \eta , \eta'  \right) = g_{ab}(\eta-\etap) {\rm exp } \left[ - k^2 \sigma_v^2 \left( {\rm e}^{\eta} - {\rm e}^{\eta'} \right)^2 / 2 \right]^2 \,,
\eeq  where $\sigma_v \equiv \frac{1}{3} \int d^3 q \frac{P^0 \left( q \right)}{q^2}$. The large-$k$ limit of the propagator emerges as the result of a summation, at any order in standard PT, of the so called `chain diagrams' , {\it i.e.} diagrams obtained by inserting, in all possible ways, linear PS's carrying `soft' momentum $q\ll k$ on the hard propagator line \cite{RPTb}. So, the first term in the expression at the RHS of \re{fullPS1} is completely determined (modulo some arbitrariness in the interpolation procedure), while the second one, the mode-mode coupling, is given by loop diagrams in which the linear propagator is replaced everywhere by the non-linear one discussed above. Due to practical limitations, these corrections include at most some of the 2-loop diagrams in this new expansion scheme. 

Now, if we perform our `PS-opening test', introduced at the end of Sect.~\ref{PTG} on the first term, we find contributions of the same sign, that do not cancel even in the equal time case, $\eta=\etap$. Opening, for instance, a single PS (in all possible ways) for all chain diagrams contributing to the resummed propagator gives an extra term 
\beq
- (\bk\cdot \bV0)^2 \frac{1}{2}\left(T(\eta) - T(0)\right)^2  G^{eik}_{ab} \left(k,  \eta , 0 \right)\,,
\eeq
in the large-$k$ limit, and  therefore an extra contribution 
\beq - (\bk\cdot \bV0)^2 \left(T(\eta) - T(0)\right)^2\, G^{eik}_{ac} \left(k,  \eta , 0 \right)G^{eik}_{bd} \left(k,  \eta , 0 \right)u_cu_d P^0(k)
\label{propshift}
\eeq
 to the first term in \re{fullPS1} for the equal time PS. This result can be generalized to the opening of any number of linear PS, leading to the exponentiation of the $- (\bk\cdot \bV0)^2 \left(T(\eta) - T(0)\right)^2$ term. Now, positive contributions do actually come from opening PS's in the mode-mode coupling term at the second line of eq.~\re{fullPS1}\footnote{These contributions are those in which the flows of time from the two sides of the opened PS point towards the two different extrema at $\eta$ and $\etap$ of \re{fullPS1}. PS of this kind are typically a few, compared to the infinite PS that can be obeyed in a resummed propagator.}, but they are not able to cancel
 the  GT-induced term from the first line, such as eq.~\re{propshift}, as long as the mode-mode coupling therm is computed at a finite order in the new expansion scheme. Analogous considerations apply to the multi point propagator expansion \cite{Bernardeau:2008fa}. 
 
The galilean invariance of nonperturbative methods based on the solution of evolution equations, such as \cite{TaruyaIII} and \cite{Anselmi:2012cn}, requires an analysis of the RHS's of the differential equations. For instance, in the scheme proposed in \cite{Anselmi:2012cn}, the RHS of the time differential equation is GT invariant both in the $k\to 0$ and in the $k \to \infty$ limit. However, the approximations done to treat the intermediate $k$ range are not manifestly GT invariant.

The numerical effect induced by mistreating galilean invariance should be carefully investigated case by case in each of these approaches, as each of them is based on a different reorganization of the perturbative expansion and on different truncation schemes. A naive estimate, obtained by computing the effect on the 1-loop PS of neglecting the contribution to the mode-mode coupling term coming from the IR modes responsible for the cancellation gives (at $z=0$ and for a $\Lambda$CDM PS) an effect around the percent at the upper extreme of the BAO range of $k$ and growing as $k^2$ at higher $k$'s. Although this procedure likely overestimates the true effect, as the 1-loop IR cancellation is accounted for completely in all the methods mentioned above, we conclude that the galilean invariance of resummation schemes clearly deserves a careful consideration, especially if one aims to extend the range of validity of a given computational scheme beyond the BAO scales.

In the remaining part of this section we discuss a resummation method introduced in \cite{Anselmi:2012cn}, and show that is galilean invariant. To our knowledge, it is the only computational scheme to pass the PS-opening test (and respecting Ward identities) among those discussed in the literature so far. A crucial role is played by counter terms, which should be properly taken into account at each order of the new expansion scheme to avoid over counting of the PT contributions. As we will see, they are also responsible for the fulfilling of the Ward identities derived in Sect.~\ref{Ward}, and therefore, guarantee the galilean invariance of the expansion.

The scheme can be considered as a variant of the RPT of  \cite{RPTa,RPTb}. The starting point is to define as the new tree-level propagator and power spectrum the quantities 
\beqra
&&G_{ab}^{\rm eik} \left( k ; \eta, \eta' \right)  \equiv  g_{ab} \left( \eta , \eta' \right) {\rm exp} \left[ - k^2 \sigma_v^2 \frac{\left( {\rm e}^\eta - {\rm e}^{\eta'} \right)^2}{2} \right]\,,\nonumber\\
&&\;\;\;\;\;\;\;\;\;\;\;\;\;\;\;\;\;\;\;= \big[ g^{-1}-  \Sigma^{eik}  \big]_{ab}^{-1}(k;\eta,\etap)\,,\nonumber\\
&&\;\;\;\;\;\;\;\;\;\;\;\;\;\;\;\;\;\;\;= g_{ab}(\eta,\etap) + \int d s ds'\,  g_{ac}(\eta,s)\Sigma^{eik}_{cd}(k;s,s')  G^{eik}_{db}(k;s',\etap)\,,
\label{Geik}
\eeqra
and
\beqra
&&P_{ab}^{\rm eik} \left( k ; \eta, \eta' \right) \equiv  P^0 \left( k \right) \, u_a u_b  {\rm exp} \left[ - k^2 \sigma_v^2 \frac{\left( {\rm e}^\eta - {\rm e}^{\eta'} \right)^2}{2} \right] \,,\nonumber\\
&&\;\;\;\;\;\;\;\;\;\;\;\;\;\;\;\;\;\;\;= G^{eik}_{ac}(k;\eta,\etain) G^{eik}_{bd}(k;\etap,\etain) P^0(k)u_a u_b \nonumber\\
&&\;\;\;\;\;\;\;\;\;\;\;\;\;\;\;\;\;\;\;+\int d s ds'\,  G^{eik}_{ac}(k;\eta,s) G^{eik}_{bd}(k;\etap,s') \Phi^{eik}_{cd}(k;s,s')\,,
\eeqra
for all values of $k$. These quantities are obtained by starting from the corresponding linear ones in standard PT, and  by adding to them  all possible chain diagrams,  in which ``soft'' power spectra of momenta $q_i$ (namely, with $q_i \ll k$) are added to the hard line of momentum $k$. The approximation in which only these contributions are taken into account at the fully non-linear level is also referred to as the ``eikonal'' approximation \cite{Bernardeau:2012aq}, due to its analogy with the resummation of soft gluon contributions in QCD. Therefore, we will refer  to these tree level quantities as the eikonal propagator and PS, respectively, and, following \cite{Anselmi:2012cn}, we will indicate the new scheme as ``eRPT''.

To take these contributions into account,  one can  \cite{Anselmi:2012cn} add and subtract the quadratic expression
\begin{eqnarray}
& &  \!\!\!\!\!\!\!\! \!\!\!\!\!\!\!\! \!\!\!\!\!\!\!\! \!\!\!\!\!\!\!\! 
i S_e \left[  \Phi_{ab}^{\rm eik} \left( k ; \eta , \eta' \right) ,  \Sigma_{ab}^{\rm eik} \left( k ; \eta , \eta' \right) \right]  \equiv  \nonumber\\
& & \!\!\!\!\!\!\!\! \!\!\!\!\!\!\!\! \!\!\!\!\!\!\!\! \!\!\!\!\!\!\!\!  \!\!\!\! 
\int d \eta d \eta' d^3 \bk \left[ - \frac{1}{2} \chi_a \left( - \bk , \eta \right) \Phi_{ab}^{\rm eik} \left( k ; \eta , \eta' \right) \chi_b \left( \bk , \eta' \right) - i  \chi_a \left( - \bk , \eta \right) \Sigma_{ab}^{\rm eik} \left( k ; \eta , \eta' \right) \varphi_b \left( \bk , \eta' \right) \right] ,\nonumber\\
\label{counter-terms}
\end{eqnarray}
in the exponent of the generating functional \re{Z}. The added term is included in the new ``free'' action, in such a way that, after after integrating out the $\vp_a$ and $\chi_a$ fields, the new free-path integral gives.~\footnote{Integrating $\varphi_a$ out, now gives $\chi_a \left( \bk , \eta \right) = - \int d \eta' J_b \left( \bk , \eta' \right) G_{ba}^{\rm eik} \left( k ; \eta' , \eta \right)$; however, one does not need to change the frame fixing term of eq.~\re{Sgf}, since $\Sigma_{ab}^{\rm eik} \left( k \right)$ vanishes at $k=0$.} 
\begin{eqnarray}
& & 
\!\!\!\!\!\!\!\!\!\!\!\!\!\!\!\! \!\!\!\!\!\!\!\!\!\!\!\!\!\!\!\! 
Z_0^{eRPT}[J_a,\, K_b;\bV0] =  {\rm exp} \Bigg\{ - \int d^3 \bk d \eta d \etap \Bigg[ \frac{1}{2} J_a \left( \bk , \eta \right) P_{ab}^{\rm eik} \left( k ; \eta , \etap \right) J_b \left( - \bf k , \etap \right)  \nonumber\\
&& \quad\quad\quad\quad\quad\quad\quad\quad
  + i J_a \left( \bk , \eta \right) G_{ab}^{\rm eik} \left( k ; \eta , \etap \right) K_b \left( - \bk , \etap \right) 
 \Bigg] \nonumber\\
 && \quad\quad  \quad\quad  \quad\quad 
   + i \int d^3 \bk d \eta J_2 \left( - \bk , \eta \right) {\bar \varphi}_2 \left( \bk , \eta , \bV0 \right)
     \Bigg\} \, .
     \label{zfree}
\end{eqnarray}

The functions $ \Phi_{ab}^{\rm eik} \left( k \right) $ and $ \Sigma_{ab}^{\rm eik} \left( k \right) $ can be represented by diagrams in which the hard line, carrying momentum of order $k$, is corrected by attaching to it soft power spectra in all possible ways so  that the final diagram is 1PI  \cite{Anselmi:2012cn}. The subtracted term, that is eq.~\re{counter-terms} with  flipped  sign, is instead included in the new interaction term. The interaction lagrangian therefore consists of the cubic term from \re{action}, plus the two counterterms  $ - \Phi_{ab}^{\rm eik} \left( k \right) $ (connecting two $\chi$)   and  $ i  \Sigma_{ab}^{\rm eik} \left( k \right) $  (connecting one $\varphi$ and one $\chi$). These counter terms avoid overcounting of the contributions already included in   \re{zfree}. By construction, eRPT and standard PT are equivalent at the infinite loop order.

Repeating the same steps leading to \re{id2} with the new splitting between free and interacting action, the generating functional can be expressed as 
\begin{eqnarray}
& & \!\!\!\!\!\!\!\! \!\!\!\!\!\!\!\! \!\!\!\!\!\!\!\! \!\!\!\!\!\!\!\! \!\!\!\!\!\!\!\!
Z \left[ J_a,  K_b;  \bV0 \right] = 
{\rm e}^{i \int d \eta d^3 \bk J_2 \left( - \bk , \eta \right) {\bar \varphi}_2 \left( \bk, \eta , \bV0 \right)}{\rm e}^{\int d \eta d^3 \bk \frac{\delta^2}{\delta K_a \left( \bk ,\, \eta \right) \delta J_a \left( - \bk ,\, \eta \right) } \bk \cdot \bV0 
\partial_\eta T \left( \eta \right) }  \times
 \nonumber\\
&&{\rm e}^{\frac{1}{2} \,\int d\eta\etap d^3 \bk  e^{-i \bk\cdot\bV0 (T(\eta)-T(\etap) )} \Bigg[ \Phi_{ab}^{\rm eik}  \left( k ; \eta , \eta' \right) \frac{-i \delta}{\delta K_a \left( - \bk , \eta \right)} \frac{-i \delta}{\delta K_b \left(  \bk , \etap \right)}    
 + i  \Sigma_{ab}^{\rm eik} \left( k ; \eta , \etap \right)
\frac{-i \delta}{\delta K_a \left( - \bk , \eta \right)} \frac{-i \delta}{\delta J_b \left(  \bk , \etap \right)} \Bigg]}\times \nonumber\\
&& {\rm e}^{ \int d \eta d^3 \bk d^3 \bq d^3 \bp \,\gamma_{abc}(\bk,\bq,\bp)\frac{\delta^3\;\;\;}{\delta K_a(\bk,\eta)\delta J_b(\bq,\eta)\delta J_c(\bp,\eta)}}  Z _0^{\rm eik}\left[ J_a,  K_b;{\bf 0 } \right]\,. \nonumber\\
\end{eqnarray} 
Setting $\bV0=0$ we get the new expansion in the eRPT scheme in the c.o.m frame. In a generic frame, comparing with  \re{id2}, we see that there are new $\bV0\neq 0$-dependent terms, coming from the transformation of the counter terms. Therefore, a GT is realized by inserting, besides the tadpole term discussed in Sect.~\ref{PTG}, the new contributions from the counter terms.

This expression indicates how a correlator transforms under a boost of velocity $\bV0$. We note that the transformation is realized by a series of mass insertions (identical to those in standard PT) and by  phases on the counterterms.

As an example, let us evaluate the GT of the propagator \re{Geik}
up to first order in $\bV0$. We obtain
\begin{eqnarray}
& &  \!\!\!\!\!\!\!\! \!\!\!\!\!\!\!\! \!\!\!\!\!\!\!\!  
G_{ab}^{\rm eik} \left( k ; \eta , \etap \right) \rightarrow G_{ab}^{\rm eik} \left( k ; \eta , \etap \right) - i \bk \cdot \bV0
\int_{\eta}^{\etap} d s \partial_s T(s)  G_{ac}^{\rm eik} \left( k ; \eta , s \right)  G_{cb}^{\rm eik} \left( k ; s , \etap \right) \nonumber\\
& &  \!\!\!\!\!\!\!\! \!\!\!\!\!\!\!\! \!\!\!\!\!\!\!\!  
- i \bk \cdot \bV0 \int_{\etap}^{\eta} d s  \int_{\etap}^{s} d s' \;
 G_{ac}^{\rm eik} \left( k ; \eta , s \right)   \Sigma_{cd}^{\rm eik} \left( k ; s , s' \right)     G_{db}^{\rm eik} \left( k ; s' , \etap \right)\;\left[ T \left( s \right) - T \left( s' \right) \right]  \,.  \nonumber\\
\label{tres1}
\end{eqnarray}
Notice that for the eikonal propagators the composition property for linear propagators, eq.~\re{composition}, does not hold, and therefore we cannot take the product of propagators out of the integral on $s$ at the first line, as we did in the PT case (see eq.~\re{P-GT-01}). Here is where the role of counter terms in the second line is crucial.
Indeed, we can write the second term at the RHS of the first line as
\beqra
&& \!\!\!\!\!\!\!\! \!\!\!\!\!\!\!\! \!\!\!\!\!\!\!\!\!\!\!\!\!\!\!\!  - i \bk \cdot \bV0
\int_{\eta}^{\etap} d s \, \partial_s T(s) G_{ac}^{\rm eik} \left( k ; \eta , s \right)  G_{cb}^{\rm eik} \left( k ; s , \etap \right) \nonumber\\
&& \!\!\!\!\!\!\!\! \!\!\!\!\!\!\!\! \!\!\!\!\!\!\!\! \!\!\!\! \!\!\!\!=  i \bk \cdot \bV0
\int_{\etap}^{\eta} d s  \int_{\etap}^{s} d s' \;G_{ac}^{\rm eik} \left( k ; \eta , s \right) g_{cd}^{-1}( s, s')  G_{db}^{\rm eik} \left( k ; s' , \etap \right) \, \left[ T \left( s \right) - T \left( s' \right) \right]  \,,
\eeqra
where $g_{cd}^{-1}( s, s') = \delta_D(s-s') \left(\delta_{cd}\partial_{s'} + \Omega_{cd}\right)$.
Summing this expression with the second line of  \re{tres1} gives
\beqra
&& \!\!\!\!\!\!\!\! \!\!\!\!\!\!\!\! \!\!\!\!\!\!\!\! \!\!\!\! \!\!\!\!=  i \bk \cdot \bV0
\int_{\etap}^{\eta} d s  \int_{\etap}^{s} d s' \;G_{ac}^{\rm eik} \left( k ; \eta , s \right) {G_{cd}^{{\rm eik}}}^{-1}(k; s, s')  G_{db}^{\rm eik} \left( k ; s' , \etap \right) \, \left[ T \left( s \right) - T \left( s' \right) \right]  \nonumber\\
&& \!\!\!\!\!\!\!\! \!\!\!\!\!\!\!\! \!\!\!\!\!\!\!\! \!\!\!\! \!\!\!\!=  i \bk \cdot \bV0
\int_{\etap}^{\eta} d s  \int_{\etap}^{s} d s' \;\big[ G_{ac}^{\rm eik} \left( k ; \eta , s \right)  \delta_{cb}   \delta_D(s-\etap) T(s)   \nonumber\\
&& \;\;\;\;\;\;\;\;\;\;\;\;\;\;\;\;\;\;\;- \delta_{ad} \delta_D(\eta-s')T(s') G_{db}^{\rm eik} \left( k ; s' , \etap \right) \big]\,,\nonumber\\
&& \!\!\!\!\!\!\!\! \!\!\!\!\!\!\!\! \!\!\!\!\!\!\!\! \!\!\!\! \!\!\!\!=- i \bk \cdot \bV0 \;\left[ T \left( \eta \right) - T \left( \etap \right) \right]\; G_{ab}^{\rm eik} \left( k ; \eta , \etap \right) \,,
\label{miracle}
\eeqra
which is exactly  the linear term in the GT transformation for a propagator, according to the general rule of eq.~\re{general-corr-GT}.

Proceeding in close analogy with the discussion for standard PT in Sect.~\ref{PTG}, we can show that a cancellation between the $\bV0$-dependent contributions takes place at any order in the eRPT expansion of the equal time PS, provided the counter terms are properly taken into account.

 We now discuss Ward identities in the eRPT scheme.  The starting formal relations, eq.~\re{dZ2}, as well as the derived ones as, for instance, eq.~\re{G2G3-full}, are unchanged, since the generating functional from which they are derived is the same for PT and for eRPT. What is modified, on the other hand, is the way in which the GI  enforced by Ward identities is implemented order by order. As in the calculation discussed above, the role of counter terms is crucial also in this context. 

For definitiveness, we consider again the Ward identity of  eq.~\re{G2G3-full}. At the tree level, the identity takes exactly the same form as for PT. Indeed the first term  now gets two contributions, one from the inverse eikonal propagator and one from the counter term, whose sum gives back the inverse linear propagator $g_{cb}^{-1}$,
\beqra
\frac{\delta^2 \Gamma^{\rm eRPT}_{\rm   tree}}{\delta \chi_c \left( \bp , \eta'' \right) \delta \varphi_b \left( \bq , \eta' \right) } \vert_{\varphi_a = \chi_b = 0}  &=&\delta_D \left( \bp + \bq \right){ G^{eik}_{cb}}^{-1}(q; \eta'',\etap)  
\nonumber\\
&&+\delta_D \left( \bp + \bq \right) \Sigma^{eik}_{cb}(q; \eta'',\etap)\,,  \nonumber\\
&=&\frac{\delta^2 \Gamma_{\rm   tree}}{\delta \chi_c \left( \bp , \eta'' \right) \delta \varphi_b \left( \bq , \eta' \right) } \vert_{\varphi_a = \chi_b = 0} \,,
\eeqra
while, since the  tree-level effective action contains just one trilinear term, we get
\beq
\frac{\delta^3 \Gamma^{\rm eRPT}_{\rm tree}}{\delta \chi_c \left( \bp , \eta'' \right) \delta \varphi_2 \left( \bk , \eta' \right) 
 \delta \varphi_b \left( \bq , \eta' \right) }=\frac{\delta^3 \Gamma_{\rm tree}}{\delta \chi_c \left( \bp , \eta'' \right) \delta \varphi_2 \left( \bk , \eta' \right) 
 \delta \varphi_b \left( \bq , \eta' \right) } \,.
\eeq
The one-loop level is less trivial. The next-to-lowest contributions to the second line of eq.~\re{G2G3-full} are given by diagrams such as those at the RHS of  Fig.~\ref{fig:w23-1loop} with the linear PS and propagator replaced by their full -eikonal- counterparts, while the vertices are kept at the tree level. As it was discussed after eq.~\re{tres1}, the two eikonal propagators (and, similarly, the eikonal propagator and the eikonal PS) meeting at the uppermost vertices do not combine into a single one, so that the LHS in Fig.~\ref{fig:w23-1loop} (in which  the linear PS and propagator  are also replaced by their full -eikonal- counterparts) is not reproduced. This is achieved, by the same mechanism of eq.~\re{miracle}, by adding to the LHS the three diagrams obtained by inserting a $\Sigma^{eik}_{ab}$ counter term in all possible ways, namely, at the left and at the right of the PS box and at the lower propagator. This mechanism works at higher orders, including higher loops and also the $\Phi^{eik}_{ab}$ counter term.

By similar arguments one can show that the  eRPT scheme fulfills all the Ward identities imposed by galilean invariance, and is therefore safe from the point of view of spurious dependence on long wavelength velocity perturbations.

\section{Conclusions}
\label{conclus}
Galilean invariance is equivalent to the statement that long wavelength velocity perturbations completely decouple from short wavelength modes in the infinite wavelength limit. This is to be contrasted with the effect of long wavelength density (or velocity {\em divergence}) perturbations, which can be seen as a modification (renormalization) of the background evolution for the short modes, as discussed in \cite{Baldauf:2011bh}. Therefore, computing a given correlator in a frame different from the c.o.m. one, in which the average velocity of particles is $\bV0\neq 0$ (in practice, by considering the tadpole contributions of Sect.~\ref{PTG}) should give the same result as the one obtained by `gauging away' $\bV0$ by a GT (and then considering the phase factors of eq.~\ref{general-corr-GT}). This line of thinking leads to the Ward identities obtained in this paper, and is the counterpart in the LSS context of what is usually done for gauge symmetries in quantum field theory of for conformal invariance in inflation, where the used of Ward identities and consistency relations is well established
\cite{Maldacena:2002vr,Creminelli:2004yq,Cheung:2007sv,Assassi:2012zq,Creminelli:2012ed,Hinterbichler:2012nm,Schalm:2012pi}.

Nonlinear relations such as eq.~\re{P-B-finalid} can be of use in many respects. From the computational point of view, as discussed in Sect. \ref{resummations}, they provide consistency checks for nonperturbative approaches aiming at extending the range of validity of PT towards smaller scales. In this paper we have identified a well defined expansion scheme, eRPT,  which, thanks to the role of counter terms fully passes the test. The numerical feasibility and performances of eRPT, as well as the implementation of  explicitly GT invariant approximation schemes to the approach in \cite{Anselmi:2012cn}, will be analyzed in a future work \cite{futureeRPT}. 

We stress again that, since these relations descend merely from the GI of the DM fluid, they hold even at scales in which the single stream approximation is broken, and therefore apply also to more refined schemes, in which velocity dispersion is taken into account, such as \cite{Pietroni:2011iz}, and to the construction of effective field theories, as in \cite{Carrasco:2012cv}.

Last, but not least, the relation between the nonlinear PS and the nonlinear bispectrum in the squeezed limit, expressed  in eq.~\re{P-B-finalid} should be eventually probed in numerical simulations, and in real observations, once large redshift surveys will be completed and the correlators at different times will be measured. The different time behavior of the two dominating terms in~\re{P-B-finalid} can possibly allow to extract the contribution from the primordial nongaussianity, as opposed to that induced by the nonlinear dynamics during the formation of the structures.  A quantitative study of this, and of other relations that follow from GI (for instance, between higher point correlators, or between correlators including the velocity field)  is definitely an interesting topic for future research.

\section*{Acknowledgments}

We thank Stefano Anselmi, Nicola Bartolo, Sabino Matarrese,  and Antonio Riotto for valuable discussions. M. Pietroni acknowledges partial support from the  European Union FP7  ITN INVISIBLES (Marie Curie Actions, PITN- GA-2011- 289442). M. Peloso  acknowledges partial support from the DOE grant DE-FG02-94ER-40823 at the University of Minnesota.   M. Peloso would like to thank the University of Padova, and INFN, Sezione di Padova, for their friendly hospitality and for partial support during his sabbatical leave.

 \section*{References}
\bibliographystyle{JHEP}
\bibliography{/Users/pietroni/Bibliografia/mybib.bib}
\end{document}